\documentclass[aps,prb,twocolumn,superscriptaddress]{revtex4-1}
\usepackage{amsmath,amssymb,dcolumn,bm,epsfig,graphicx,longtable}
\begin{document}

\title{Nonlocal van der Waals functionals for solids: Choosing an appropriate one}
\author{Fabien Tran}
\affiliation{Institute of Materials Chemistry, Vienna University of Technology,
Getreidemarkt 9/165-TC, A-1060 Vienna, Austria}
\author{Leila Kalantari}
\affiliation{Institute of Materials Chemistry, Vienna University of Technology,
Getreidemarkt 9/165-TC, A-1060 Vienna, Austria}
\author{Boubacar Traor\'{e}}
\affiliation{Univ Rennes, INSA Rennes, CNRS, Institut FOTON - UMR 6082, F-35000 Rennes, France}
\author{Xavier Rocquefelte}
\affiliation{Univ Rennes, ENSCR, INSA Rennes, CNRS, ISCR (Institut des Sciences Chimiques
de Rennes) - UMR 6226, F-35000 Rennes, France}
\author{Peter Blaha}
\affiliation{Institute of Materials Chemistry, Vienna University of Technology,
Getreidemarkt 9/165-TC, A-1060 Vienna, Austria}

\begin{abstract}

The nonlocal van der Waals (NL-vdW) functionals [Dion \textit{et al}.,
Phys. Rev. Lett. \textbf{92}, 246401 (2004)] are being applied more and more
frequently in solid-state physics, since they have shown to be much more reliable
than the traditional semilocal functionals for systems where weak
interactions play a major role. However, a certain number of
NL-vdW functionals have been proposed during the last few years, such that
it is not always clear which one should be used.
In this work, an assessment of NL-vdW functionals is presented.
Our test set consists of weakly bound solids, namely rare gases,
layered systems like graphite, and molecular solids, but also strongly bound
solids in order to provide a more general conclusion about the accuracy
of NL-vdW functionals for extended systems. We found that among the
tested functionals, rev-vdW-DF2 [Hamada, Phys. Rev. B \textbf{89}, 121103(R) (2014)]
is very accurate for weakly bound solids, but also quite reliable for strongly bound solids.

\end{abstract}

\maketitle

\section{\label{introduction}Introduction}

The pioneering works of Langreth, Lundqvist, and co-workers on nonlocal van der
Waals functionals\cite{RydbergPRL03,DionPRL04,ThonhauserPRB07,LangrethJPCM09,HyldgaardPRB14,BerlandRPP15}
(abbreviated as NL-vdW) have contributed significantly in making density
functional theory \cite{HohenbergPR64,KohnPR65} (DFT) much more accurate
for extended systems where the weak vdW interactions are important.
Before the advent of the NL-vdW functionals and atom-pairwise methods
\cite{BeckeJCP05,TkatchenkoPRL09,GrimmeJCP10}
(see Refs.~\onlinecite{GrimmeCR16,HermannCR17} for reviews),
the DFT calculations with periodic boundary conditions were done mostly with
the local density approximation (LDA)\cite{KohnPR65} or generalized gradient approximation
(GGA).\cite{PerdewPRL96} However, since the physics of the London dispersion
forces is included neither in LDA and GGA nor in meta-GGA (MGGA) and
hybrid functionals, all these methods are in general quite unreliable for the
calculation of the length and binding energy of noncovalent bonds.

Thus, the NL-vdW methods, the focus of this work, are becoming increasingly
popular, and particularly in the solid-state community\cite{ChoudharyPRB18} thanks to the
availability of computationally fast implementations.\cite{RomanPerezPRL09,SabatiniJPCM12,WuJCP12,LarsenMSMSE17,TranPRB17}
However, since there is still no full confidence in the accuracy of the results
when using a NL-vdW functional (the so-called chemical accuracy of 1~kcal/mol can sometimes be
reached, but not systematically, see e.g. Refs.~\onlinecite{KlimesJPCM10,LobodaJCP18}),
new ones are constantly being proposed and currently more than twenty exist
(see, e.g., Refs.~\onlinecite{KlimesPRB11,BjorkmanPRB12,BerlandPRB14,PengPRB17}).
Consequently, as in the case of semilocal (i.e., GGA and MGGA) and hybrid functionals,
there is a rather large freedom in the choice of the NL-vdW functional and
it may not be always clear which one to choose.

In our previous work,\cite{TranJCP16} a plethora of functionals of the first four
rungs of DFT Jacob's ladder\cite{PerdewAIP01} were tested on a set consisting of strongly bound
and weakly bound solids. Functionals including a term of the atom-pairwise
type to account for the vdW interactions,
DFT+D3 (Ref.~\onlinecite{GrimmeJCP10}) and DFT+D3(BJ) (Ref.~\onlinecite{GrimmeJCC11}),
were also considered. Here, we extend this comparison by considering
NL-vdW functionals, not considered in Ref.~\onlinecite{TranJCP16},
and the main goal is to provide a useful summary of their performance
for the geometry and binding energy of periodic solids.

Additionally, results for the binding energy of molecules will also be shown in
order to provide a hint on the performance of the tested functionals on finite systems.

The paper is organized as follows. Section~\ref{methods} gives
details about the methods. Then, the results are presented and discussed in
Sec.~\ref{results} and summarized in Sec.~\ref{summary}.

\section{\label{methods}Methods}

In NL-vdW methods,\cite{DionPRL04} the exchange-correlation (xc) functional
is given by
\begin{equation}
E_{\text{xc}} = E_{\text{xc}}^{\text{SL/hybrid}} +
E_{\text{c,disp}}^{\text{NL}},
\label{Exc}
\end{equation}
where the first term is of the semilocal (SL) or
hybrid\cite{BeckeJCP93a,BeckeJCP93b} type and
the second term reads
\begin{equation}
E_{\text{c,disp}}^{\text{NL}} = \frac{1}{2}\int\int\rho(\bm{r}_{1})
\Phi\left(\bm{r}_{1},\bm{r}_{2}\right)\rho(\bm{r}_{2})d^{3}r_{1}d^{3}r_{2}.
\label{EcdispNL}
\end{equation}
Thanks to its form, the additional correlation term given by
Eq.~(\ref{EcdispNL}) is able to account for long-range interactions in the
system; the specificity of $E_{\text{c,disp}}^{\text{NL}}$ is to
contribute to the binding energy between two systems A and B even when there is
no density overlap
[i.e., $\rho_{\text{A}}(\bm{r})\rho_{\text{B}}(\bm{r})=0$ $\forall \bm{r}$],
while in such a case the contribution from $E_{\text{xc}}^{\text{SL/hybrid}}$ is
strictly zero (if we neglect the change in the shape of $\rho_{\text{A}}$ and
$\rho_{\text{B}}$ when the system A$-$B is formed).
The kernel $\Phi$ in Eq.~(\ref{EcdispNL}) depends on the electron
density $\rho$, its derivative $\nabla\rho$, and the interelectronic distance
$\left\vert\bm{r}_{1}-\bm{r}_{2}\right\vert$. To our knowledge, five different
analytical forms for $\Phi$ have been proposed to date,
\cite{DionPRL04,VydrovPRL09,VydrovJCP10,SabatiniPRB13,TerentjevPRB18b}
while numerous reoptimizations of the parameters in $\Phi$
have been reported.\cite{LeePRB10,BjorkmanPRB12,AragoJCTC13,MardirossianPCCP14,PengPRX16,PengPRB17}

The choice of the semilocal or hybrid functional
$E_{\text{xc}}^{\text{SL/hybrid}}$ in Eq.~(\ref{Exc}) is also of crucial
importance, since this is of course the total xc-functional
which has to be accurate. In particular, an important requirement is that
$E_{\text{xc}}^{\text{SL/hybrid}}$ alone should not already lead to an overbinding,
otherwise adding $E_{\text{c,disp}}^{\text{NL}}$
can only make the results worse. Thus, the combination
$E_{\text{xc}}^{\text{SL/hybrid}}+E_{\text{c,disp}}^{\text{NL}}$ has to be
well-balanced in order to provide accurate geometry and binding
energy.\cite{LeePRB10,KlimesPRB11,BerlandPRB14,HamadaPRB14}

Among the NL-vdW functionals that are available in the literature,
a certain number of them were selected for the present work.
However, we did not consider NL-vdW functionals based on hybrid functionals,
\cite{AragoJCTC13,MardirossianPCCP14,BerlandJCP17,JiaoJCP18}
since they lead to calculations that are much more expensive,
especially for solids. They are therefore less interesting from
a practical point of view as long as the electronic structure
is of no particular interest. Thus, only semilocal-based NL-vdW functionals
are considered and now listed.

vdW-DF from Dion \textit{et al}. (DRSLL),\cite{DionPRL04} the first proposed
NL-vdW functional that can be applied to systems with arbitrary geometry,
consists of the GGA exchange revPBE\cite{ZhangPRL98} (a reoptimization of
PBE\cite{PerdewPRL96}) and LDA correlation\cite{VoskoCJP80,PerdewPRB92a} for
the semilocal part. The nonlocal term, Eq.~(\ref{EcdispNL}),
of the vdW-DF functional (DRSLL kernel $\Phi$) has
subsequently been used in combination with other semilocal components,
and among them, those that are considered in the present work are the following
four. C09-vdW from Cooper,\cite{CooperPRB10} which uses a GGA (C09x)
for exchange and LDA for correlation.
optB88-vdW and optB86b-vdW, which are two of the functionals developed by
Klime\v{s} \textit{et al}.,\cite{KlimesJPCM10,KlimesPRB11}
and use for the semilocal component, the GGAs optB88 and optB86b
for exchange and LDA for correlation.
Note that optB88 and optB86b are reoptimizations of
B88\cite{BeckePRA88} and B86b,\cite{BeckeJCP86b} respectively.
vdW-DF-cx from Berland and Hyldgaard,\cite{BerlandPRB14} which consists of
a GGA exchange component, LV-PW86r, that is combined with LDA correlation
and was constructed to be more consistent with the DRSLL kernel.

vdW-DF2 from Lee \textit{et al}. (LMKLL)\cite{LeePRB10}
uses the GGA exchange PW86R\cite{LeePRB10}
(a reoptimization of PW86\cite{PerdewPRB86}) and LDA for correlation, while
the kernel $\Phi$ (called LMKLL) in $E_{\text{c,disp}}^{\text{NL}}$
has the same analytical form as the original DRSLL kernel, but with
a reoptimized parameter $Z_{ab}$ ($Z_{ab}=-0.8491$ in DRSLL and
$Z_{ab}=-1.887$ in LMKLL). Hamada\cite{HamadaPRB14} proposed a revised
vdW-DF2 (rev-vdW-DF2) which combines the GGA B86R for exchange
(another reoptimization of B86b\cite{BeckeJCP86b}) and LDA correlation
with the LMKLL kernel.

Based on a kernel $\Phi$ that has a different analytical form,
rVV10\cite{SabatiniPRB13} consists of PW86R (exchange) and PBE
(correlation) for the semilocal component.
Note that the rVV10 kernel is based on the VV10 kernel of Vydrov and
Van Voorhis\cite{VydrovJCP10} and was made suitable for the method of
Rom\'{a}n-P\'{e}rez and Soler\cite{RomanPerezPRL09} (RPS) to calculate
Eq.~(\ref{EcdispNL}).
Also tested are SCAN+rVV10 and PBE+rVV10L from
Peng \textit{et al}.,\cite{PengPRX16,PengPRB17} where the MGGA
SCAN\cite{SunPRL15} and GGA PBE\cite{PerdewPRL96} are supplemented by the
NL-vdW term rVV10, but with reoptimizations of the parameter $b$
($b=6.3$, 15.7, and 10 in rVV10, SCAN+rVV10, and PBE+rVV10L, respectively).
Note that the rVV10 kernel contains another parameter, $C$, whose original value
(0.0093) was kept in SCAN+rVV10 and PBE+rVV10L.

Finally, the PBEsol+rVV10s functional proposed very recently by
Terentjev \textit{et al}.\cite{TerentjevPRB18b} will also be considered.
PBEsol+rVV10s uses the PBEsol GGA functional\cite{PerdewPRL08}
for $E_{\text{xc}}^{\text{SL}}$ and a modified rVV10 kernel (rVV10s), where
$b=10$ (as in PBE+rVV10L) and $C$ is replaced
by a function of the reduced density gradient
$s=\left\vert\nabla\rho\right\vert/\left(2\left(3\pi^{2}\right)^{1/3}\rho^{4/3}\right)$:
\begin{equation}
C(s) = C_{0} + \frac{C_{1}}{1 + C_{2}\left(s-\frac{1}{2}\right)^{2}},
\label{C}
\end{equation}
where $C_{0}=0.0093$, $C_{1}=0.5$, and $C_{2}=300$.

For the sake of comparison, results obtained with LDA, the GGAs
PBE and PBEsol, the MGGAs SCAN and TM,\cite{TaoPRL16} as well as
two atom-pairwise DFT+D3(BJ) methods [PBE-D3(BJ) and revPBE-D3(BJ)\cite{GrimmeJCC11}
including the three-body non-additive dispersion term\cite{GrimmeJCP10}]
will also be shown. SCAN and TM are modern functionals which have been shown
to be overall more accurate than GGA functionals.
\cite{TranJCP16,MoPRB17,IsaacsPRM18,MejiaRodriguezPRB18,ZhangNJP18,TranJCP18,JanaJCP18a,JanaJCP18b,ZhangNPJCM18}

The NL-vdW functionals considered here do not constitute an exhaustive list
(a few other non-hybrids can be found in
Refs.~\onlinecite{KlimesPRB11,BjorkmanPRB12,LundgaardPRB16,TerentjevC18}),
however they should represent most trends in the results that may
be obtained with this group of functionals.

\begin{figure}
\includegraphics[width=\columnwidth]{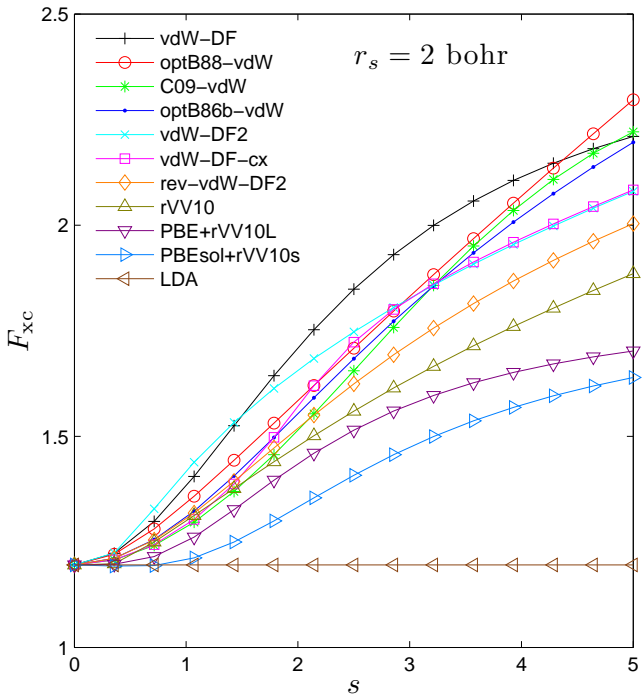}
\caption{\label{fig_Fxc}GGA Enhancement factor $F_{\text{xc}}$ of the
semilocal component of the NL-vdW functionals
plotted as a function of $s$ for $r_{s}=2$~bohr. LDA is also shown.}
\end{figure}

Figure~\ref{fig_Fxc} shows the enhancement factor of the semilocal component of the
GGA-based NL-vdW functionals, which is defined as the ratio between a GGA
xc-energy density $\epsilon_{\text{xc}}^{\text{GGA}}$ and the LDA exchange-only
$\epsilon_{\text{x}}^{\text{LDA}}$:
\begin{equation}
F_{\text{xc}}(r_{s}(\bm{r}),s(\bm{r})) =
\frac{\epsilon_{\text{xc}}^{\text{GGA}}(\bm{r})}
{\epsilon_{\text{x}}^{\text{LDA}}(\bm{r})}.
\label{Fxc}
\end{equation}
The functions $F_{\text{xc}}$ are plotted as functions of $s$ for a value of
$r_{s}=2$~bohr where $r_{s}=\left(3/\left(4\pi\rho\right)\right)^{1/3}$ is the
Wigner-Seitz radius. This comparison of the
enhancement factors will be useful later when discussing the trends in the
results, in particular for strongly bound systems where the dispersion term
does not play a major role.

\begin{figure}
\includegraphics[width=\columnwidth]{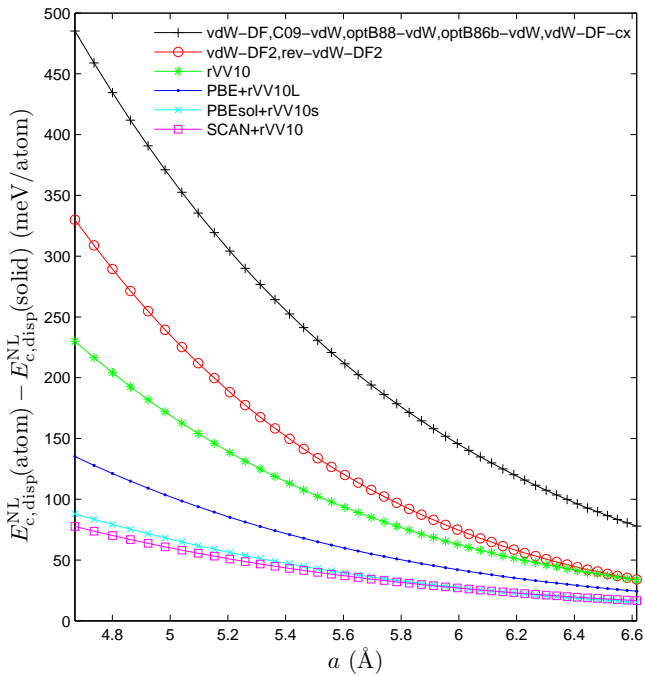}
\caption{\label{fig_kernel}Contribution to the cohesive energy of Ar coming
from the NL dispersion term plotted as a function of the lattice constant.
This is shown for all six different kernels considered in this work.}
\end{figure}

All six variants of the nonlocal dispersion term [Eq.~(\ref{EcdispNL})] considered in
this work are compared in Fig.~(\ref{fig_kernel}) which shows the
contribution $E_{\text{c,disp}}^{\text{NL}}(\text{atom})-
E_{\text{c,disp}}^{\text{NL}}(\text{solid})$ to the cohesive energy
$E_{\text{coh}}$ of solid Ar (positive values correspond to binding).
Since $E_{\text{c,disp}}^{\text{NL}}$
represents mainly the dispersion, which is attractive, the curves in
Fig.~(\ref{fig_kernel}) are positive (stronger bonding) and
have a negative slope (which favors shorter bond lengths).
It can be seen that the magnitude varies dramatically among the different
expressions for Eq.~(\ref{EcdispNL}). The original DRSLL kernel $\Phi$, that
is used in five of the functionals, leads to a contribution to
$E_{\text{coh}}$ that is the largest and one order of magnitude larger
than with the two rVV10-type kernels used in PBEsol+rVV10s and SCAN+rVV10.
Since the curve with the steepest slope is also obtained with the DRSLL kernel,
then the effect on the bond length should also be the largest when using this
kernel. We mention that the ordering of the curves observed for Ar should remain the
same for all or at least most other systems.

The calculations on periodic solids were done with the \textsc{WIEN2k} code,\cite{WIEN2k}
which is a full-potential and all-electron code based on the linearized
augmented plane-wave method.\cite{AndersenPRB75,Singh}
The implementation of the NL-vdW functionals into \textsc{WIEN2k}
has been reported recently\cite{TranPRB17} and uses
the RPS method\cite{RomanPerezPRL09} to evaluate efficiently
the NL dispersion energy [Eq.~(\ref{EcdispNL})] and the potential
$v_{\text{c,disp}}^{\text{NL}}=\delta E_{\text{c,disp}}^{\text{NL}}/\delta\rho$
entering into the Kohn-Sham equations. Since the RPS method
is based on fast Fourier transforms, it is necessary
to smooth the all-electron density $\rho$ around the nuclei, otherwise a plane-wave
expansion of $\rho$ in the whole unit cell would be practically impossible.
The smooth density that is used for the RPS method is given by
\begin{equation}
\rho_{\text{s}}(\bm{r}) =
\left\{
\begin{array}{l@{\quad}l}
\rho(\bm{r}), &
\rho(\bm{r}) \leqslant \rho_{\text{c}} \\
\frac{\rho(\bm{r})+A\rho_{\text{c}}\left(\rho(\bm{r})-\rho_{\text{c}}\right)}
{1 + A\left(\rho(\bm{r})-\rho_{\text{c}}\right)}, & \rho(\bm{r}) > \rho_{\text{c}}
\end{array},
\right.
\label{rhos}
\end{equation}
where $A=1$~bohr$^{3}$ and $\rho_{\text{c}}$ is the density cutoff
that determines the degree of smoothness applied to $\rho$.
As discussed in detail in Ref.~\onlinecite{TranPRB17}, $\rho_{\text{c}}$ has to
be chosen low enough so that the plane-wave expansion of $\rho_{\text{s}}$
is small enough to avoid too expensive fast Fourier transforms.
On the other hand, $\rho_{\text{c}}$ should also not be too low,
otherwise some accuracy with respect to the calculation with the original
density $\rho$ may be lost.

We mention that the \textsc{WIEN2k} calculations with the DRSLL and LMKLL
kernels presented in this work (but not those in our previous work\cite{TranPRB17})
were obtained with the version of these kernels generalized for spin-polarized
systems,\cite{ThonhauserPRL15} which is relevant for the calculations
of the cohesive energy\cite{GharaeePRB17}
(most atoms are spin-polarized) and for bulk Ni
which is ferromagnetic. In such spin-polarized cases, Eq.~(\ref{rhos}) is
first applied to the total density $\rho=\rho_{\uparrow}+\rho_{\downarrow}$,
then the smooth spin-$\sigma$ densities $\rho_{\text{s},\sigma}$ are obtained by
multiplying $\rho_{\sigma}$ by $\rho_{\text{s}}/\rho$:
$\rho_{\text{s},\sigma}=\rho_{\sigma}\rho_{\text{s}}/\rho$.
The calculations with the functionals of the rVV10-family were done
with the non-spin-polarized version of the kernel, since apparently no
spin-polarized version has been proposed or used
(in particular, the implementation of the rVV10 kernel in the
\textsc{Quantum ESPRESSO}\cite{SabatiniPRB13,GiannozziJPCM17}
code is non-spin polarized).

The usual parameters, like the size of the basis set or
the number of $\bm{k}$-points for the integrations in the Brillouin zone,
were chosen such that the results are well converged.
As in our previous works,\cite{TranJCP16,TranJCP18,KovacsJCP19} the results for the strongly bound
solids were obtained non-self-consistently using the PBE orbitals and density,
however the self-consistent effects
are in general quite small, below 0.005~\AA~in most cases
(the optB88-vdW results from the present work can be compared to those obtained
self-consistently in Ref.~\onlinecite{TranPRB17}).
The calculations on the weakly bound solids were done self-consistently,
except those obtained with MGGA functionals (SCAN, SCAN+rVV10, and TM)
that were done non-self-consistently using the PBE orbitals and density
since the potential of MGGA functionals is not implemented in \textsc{WIEN2k}.

\begin{table}
\caption{\label{solids}The test set of 44 strongly and 17 weakly bound solids.
The space group is indicated in parenthesis.}
\begin{ruledtabular}
\begin{tabular}{l}
Strongly bound solids \\
C ($Fd\overline{3}m$), Si ($Fd\overline{3}m$), Ge ($Fd\overline{3}m$), Sn ($Fd\overline{3}m$), \\
SiC ($F\overline{4}3m$), BN ($F\overline{4}3m$), BP ($F\overline{4}3m$), AlN ($F\overline{4}3m$), \\
AlP ($F\overline{4}3m$), AlAs ($F\overline{4}3m$), GaN ($F\overline{4}3m$), GaP($F\overline{4}3m$), \\
GaAs ($F\overline{4}3m$), InP ($F\overline{4}3m$), InAs ($F\overline{4}3m$), InSb ($F\overline{4}3m$), \\
LiH ($Fm\overline{3}m$), LiF ($Fm\overline{3}m$), LiCl ($Fm\overline{3}m$), NaF ($Fm\overline{3}m$), \\
NaCl ($Fm\overline{3}m$), MgO ($Fm\overline{3}m$), Li ($Im\overline{3}m$), Na ($Im\overline{3}m$), \\
Al ($Fm\overline{3}m$), K ($Im\overline{3}m$), Ca ($Fm\overline{3}m$), Rb ($Im\overline{3}m$), \\
Sr ($Fm\overline{3}m$), Cs ($Im\overline{3}m$), Ba ($Im\overline{3}m$), V ($Im\overline{3}m$), \\
Ni ($Fm\overline{3}m$), Cu ($Fm\overline{3}m$), Nb ($Im\overline{3}m$), Mo ($Im\overline{3}m$), \\
Rh ($Fm\overline{3}m$), Pd ($Fm\overline{3}m$), Ag ($Fm\overline{3}m$), Ta ($Im\overline{3}m$), \\
W ($Im\overline{3}m$), Ir ($Fm\overline{3}m$), Pt ($Fm\overline{3}m$), Au ($Fm\overline{3}m$) \\
\hline
Weakly bound solids \\
Rare gases: Ne ($Fm\overline{3}m$), Ar ($Fm\overline{3}m$), Kr ($Fm\overline{3}m$), \\
\hspace{0.4cm}Xe ($Fm\overline{3}m$) \\
Layered solids: graphite ($P6_{3}/mmc$), h-BN ($P6_{3}/mmc$), \\
\hspace{0.4cm}TiS$_{2}$ ($P\overline{3}m1$), TiSe$_{2}$ ($P\overline{3}m1$), MoS$_{2}$ ($P6_ {3}/mmc$), \\
\hspace{0.4cm}MoSe$_{2}$ ($P6_ {3}/mmc$), MoTe$_{2}$ ($P6_ {3}/mmc$), \\
\hspace{0.4cm}HfTe$_{2}$ ($P\overline{3}m1$), WS$_{2}$ ($P6_{3}/mmc$), WSe$_{2}$ ($P6_{3}/mmc$) \\
Molecular solids: NH$_{3}$ ($P2_{1}3$), CO$_{2}$ ($Pa\overline{3}$), \\
\hspace{0.4cm}C$_{6}$H$_{12}$N$_{4}$ ($I\overline{4}3m$)
\end{tabular}
\end{ruledtabular}
\end{table}

The list of solids composing our test set can be found in Table~\ref{solids}
along with their space group. This set consists of 44 solids with (relatively)
strong bonding of the metallic, ionic, or covalent type, and 17 solids with
weak noncovalent bonding. The strongly bound, rare-gas, and molecular solids
have a cubic cell, while the structure of the layered solids are based on the
stacking of hexagonal layers. The reference values for the lattice constants
and binding energies, to which the DFT values will be compared in Sec.~\ref{results},
were obtained either from experiment or from accurate \textit{ab initio} methods.
Most of these values, except those for the lattice constant of the
molecular solids, are corrected for the thermal and zero-point
vibrational effects, such that a direct comparison with our DFT values is possible.

As already mentioned, results for the atomization energy of
molecules will also be shown. Our test set is the AE6 set of six molecules
(SiH$_{4}$, SiO, S$_{2}$, C$_{3}$H$_{4}$, C$_{2}$H$_{2}$O$_{2}$,
C$_{4}$H$_{8}$), which were selected to give a fair idea of the accuracy of
quantum chemistry methods.\cite{LynchJPCA03} Most of the calculations were
obtained with the Gaussian augmented plane-wave method as implemented
in the \textsc{CP2K} code,\cite{VandeVondeleCPC05} which allows calculations
with NL-vdW functionals.\cite{TranJCP13} However, since not all functionals are
available in \textsc{CP2K}, calculations were also done with the
\textsc{VASP}\cite{KressePRB96} (based on the projector augmented wave method\cite{BlochlPRB94b})
and \textsc{deMon} (using Gaussian basis functions) codes.\cite{deMon}
We mention that the spin-polarized versions of the DRSLL and LMKLL
kernels are not available in the \textsc{CP2K} and \textsc{VASP} codes,
therefore the calculations were done with the spin-unpolarized form of the kernels.
However, the results were then approximately corrected by adding the spin
correction (the difference between the spin- and non-spin-polarized versions
of $E_{\text{c,disp}}^{\text{NL}}$) calculated with the \textsc{WIEN2k} code.
Furthermore, since many of our results obtained with NL-vdW functionals for the
AE6 test set strongly disagree with those presented in
Ref.~\onlinecite{CallsenPRB15}, the \textsc{VASP} results were also used to
cross-check the \textsc{CP2K} results (alternatively, \textsc{WIEN2k} could
have been used for this purpose).

To finish, we mention that \textsc{Libxc}, a library of exchange-correlation
functionals,\cite{MarquesCPC12,LehtolaSX18} has been used for some of the
calculations done with the \textsc{CP2K} and \textsc{WIEN2k} codes.

\section{\label{results}Results and discussion}

\subsection{\label{strongly}Strongly bound solids}

\begin{table*}
\caption{\label{table_strong}The ME, MAE, MRE, MARE, and MAXRE with respect to
the experimental values (corrected for thermal and
zero-point vibrational effects\cite{SchimkaJCP11,LejaeghereCRSSMS14}) on the
testing set of 44 strongly bound solids for the lattice constant $a_{0}$,
bulk modulus $B_{0}$, and cohesive energy $E_{\text{coh}}$.
The units of the ME and MAE are \AA, GPa, and eV/atom for $a_{0}$, $B_{0}$,
and $E_{\text{coh}}$, respectively, and \% for the MRE, MARE, and MAXRE.
The solid for which the MAXRE occurs is indicated in parenthesis.
All results were obtained non-self-consistently using PBE orbitals/density.
The functionals are separated into two groups, those which contain
a dispersion term (NL or atom-pairwise), and those which do not.
Within each group, the functionals are ordered by increasing
value of the MARE of $a_{0}$.}
{\tiny
\begin{ruledtabular}
\begin{tabular}{lddddcddddcddddc}
\multicolumn{1}{l}{} &
\multicolumn{5}{c}{$a_ {0}$} &
\multicolumn{5}{c}{$B_ {0}$} &
\multicolumn{5}{c}{$E_{\text{coh}}$} \\
\cline{2-6}\cline{7-11}\cline{12-16}
\multicolumn{1}{l}{Method} &
\multicolumn{1}{c}{ME} &
\multicolumn{1}{c}{MAE} &
\multicolumn{1}{c}{MRE} &
\multicolumn{1}{c}{MARE} &
\multicolumn{1}{c}{MAXRE} &
\multicolumn{1}{c}{ME} &
\multicolumn{1}{c}{MAE} &
\multicolumn{1}{c}{MRE} &
\multicolumn{1}{c}{MARE} &
\multicolumn{1}{c}{MAXRE} &
\multicolumn{1}{c}{ME} &
\multicolumn{1}{c}{MAE} &
\multicolumn{1}{c}{MRE} &
\multicolumn{1}{c}{MARE} &
\multicolumn{1}{c}{MAXRE} \\
\hline
Without dispersion           \\
TM  &    -0.006 &     0.023 &      -0.2 &       0.5 &      -1.8 (Na) &       2.4 &       6.6 &       2.1 &       6.2 &      25.2 (Cu) &      0.24 &      0.27 &       6.4 &       7.0 &      26.8 (Cu) \\
SCAN  &     0.018 &     0.030 &       0.3 &       0.6 &       3.8 (Cs) &       3.5 &       7.4 &      -0.4 &       6.5 &     -22.0 (Rb) &     -0.02 &      0.19 &      -0.7 &       4.9 &     -16.6 (Cs) \\
PBEsol  &    -0.005 &     0.030 &      -0.1 &       0.6 &      -2.3 (Sr) &       0.7 &       7.8 &      -1.4 &       7.0 &      19.5 (Ni) &      0.29 &      0.31 &       6.1 &       6.9 &      22.8 (Ni) \\
PBE  &     0.056 &     0.061 &       1.1 &       1.2 &       2.8 (Sn) &     -11.2 &      12.2 &      -9.8 &      11.0 &     -25.5 (Ge) &     -0.13 &      0.19 &      -3.9 &       5.0 &     -21.0 (Au) \\
LDA  &    -0.071 &     0.071 &      -1.5 &       1.5 &      -4.9 (Ba) &      10.1 &      11.5 &       8.1 &       9.4 &      32.8 (Ni) &      0.77 &      0.77 &      17.2 &      17.2 &      38.7 (Ni) \\
With dispersion         \\
SCAN+rVV10  &     0.004 &     0.022 &      0.0 &       0.5 &       2.5 (Cs) &       6.0 &       8.4 &       1.8 &       6.6 &      22.9 (Cu) &      0.11 &      0.22 &       2.9 &       5.4 &      17.6 (Cu) \\
PBEsol+rVV10s  &    -0.019 &     0.034 &      -0.4 &       0.7 &      -3.0 (Ba) &       3.2 &       8.1 &       0.9 &       6.3 &      22.8 (Ni) &      0.45 &      0.45 &      10.5 &      10.6 &      27.7 (Ni) \\
C09-vdW &    -0.009 &     0.037 &      -0.2 &       0.8 &      -3.2 (Ba) &       0.1 &       7.7 &      -0.9 &       6.5 &      18.2 (V) &      0.27 &      0.28 &       5.4 &       6.7 &      20.2 (Ni) \\
vdW-DF-cx &     0.015 &     0.041 &       0.3 &       0.9 &      -2.5 (Ba) &      -2.3 &       8.4 &      -4.2 &       8.0 &     -18.9 (NaF) &      0.14 &      0.19 &       2.5 &       4.8 &      16.8 (Ir) \\
PBE-D3(BJ)  &    -0.002 &     0.042 &      -0.1 &       0.9 &      -3.1 (Li) &      -3.1 &       7.5 &      -2.1 &       7.4 &     -22.6 (Ge) &      0.20 &      0.21 &       4.8 &       5.2 &      15.7 (Ni) \\
optB86b-vdW &     0.015 &     0.046 &       0.3 &       0.9 &      -2.2 (Sr) &      -5.5 &       8.2 &      -4.6 &       7.0 &     -21.5 (Ge) &      0.09 &      0.16 &       1.4 &       4.0 &      13.6 (Ir) \\
PBE+rVV10L  &     0.029 &     0.045 &       0.6 &       0.9 &       2.1 (Sn) &      -6.9 &       8.9 &      -5.9 &       7.8 &     -21.6 (Ge) &      0.10 &      0.17 &       2.4 &       4.1 &      13.1 (Ni) \\
rev-vdW-DF2 &     0.020 &     0.047 &       0.4 &       0.9 &      -2.1 (Sr) &      -6.8 &       9.0 &      -5.7 &       7.8 &     -23.6 (Ge) &      0.02 &      0.14 &      -0.9 &       4.0 &     -12.3 (Rb) \\
revPBE-D3(BJ)  &    -0.011 &     0.043 &      -0.4 &       1.0 &      -4.8 (Li) &      -0.4 &       8.5 &      -1.4 &       8.6 &     -23.2 (Ge) &      0.18 &      0.21 &       4.2 &       5.2 &      18.7 (Cu) \\
optB88-vdW &     0.026 &     0.062 &       0.6 &       1.3 &      -2.8 (Cs) &     -10.3 &      11.5 &      -6.8 &       9.2 &     -26.3 (Ge) &     -0.04 &      0.13 &      -2.0 &       3.8 &     -13.4 (Na) \\
rVV10  &     0.042 &     0.083 &       1.0 &       1.7 &       3.4 (Au) &     -13.4 &      14.5 &      -7.5 &      10.7 &     -29.8 (Ge) &      0.04 &      0.13 &       1.1 &       3.2 &      10.4 (Ba) \\
vdW-DF &     0.105 &     0.106 &       2.2 &       2.2 &       4.6 (Au) &     -23.0 &      23.2 &     -16.6 &      17.1 &     -43.0 (Au) &     -0.51 &      0.51 &     -12.8 &      12.8 &     -32.4 (Au) \\
vdW-DF2 &     0.117 &     0.140 &       2.5 &       3.0 &       6.9 (Au) &     -29.4 &      29.5 &     -18.2 &      20.4 &     -51.7 (Au) &     -0.59 &      0.59 &     -15.6 &      15.6 &     -35.4 (Sr) \\
\end{tabular}
\end{ruledtabular}
}
\end{table*}

\begin{figure}
\includegraphics[width=\columnwidth]{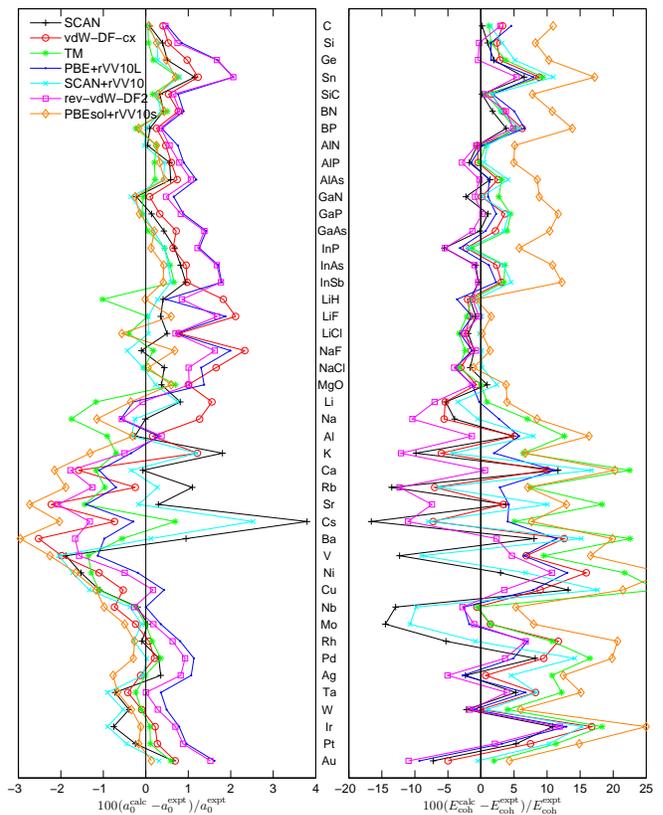}
\caption{\label{fig_solids_1}Relative error (in \%) in the calculated lattice
constant (left panel) and cohesive energy (right panel) for the 44 strongly
bound solids shown for selected functionals.}
\end{figure}

\begin{figure}
\includegraphics[width=\columnwidth]{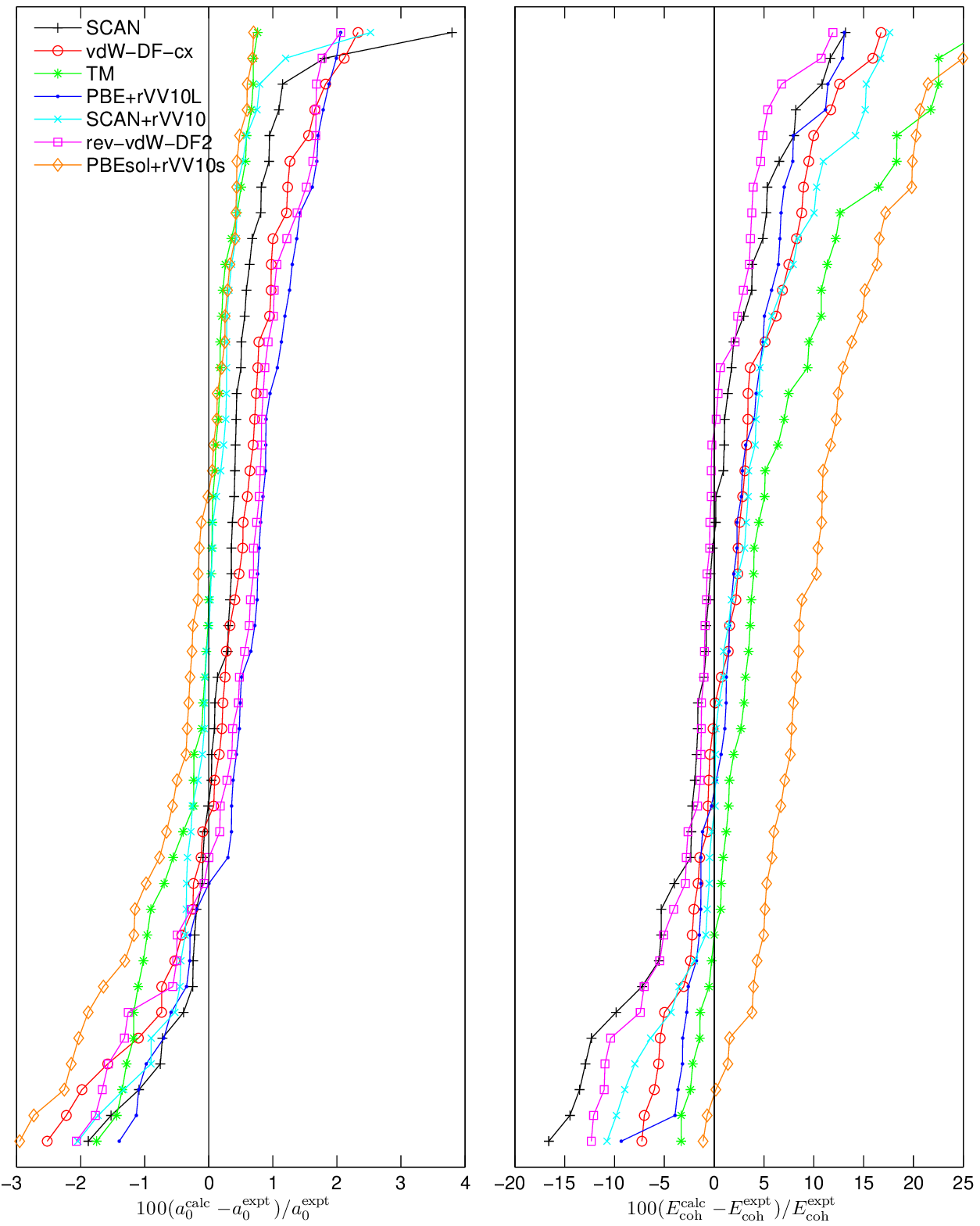}
\caption{\label{fig_solids_2}Same as Fig.~\ref{fig_solids_1}, but with the
difference that for each functional the solids have been ordered such that the relative
error goes in the direction of the positive values from bottom to top.}
\end{figure}

The results for the equilibrium lattice constant $a_{0}$, bulk modulus $B_{0}$,
and cohesive energy $E_{\text{coh}}$ of strongly bound solids obtained
with the tested xc-functionals are shown in Table~\ref{table_strong}.
ME, MAE, MRE, and MARE represent the mean values of the error, absolute error,
relative error, and absolute relative error with respect to experiment,
respectively, while MAXRE is the maximum relative error.
The experimental values were corrected for thermal and zero-point vibrational
effects.\cite{SchimkaJCP11,LejaeghereCRSSMS14}
For a few selected functionals, Figs.~\ref{fig_solids_1} and
\ref{fig_solids_2}
show graphically the errors. All detailed results are available in Tables~S1-S9 and
shown graphically in Figs.~S1-S18 of Ref.~\onlinecite{SM_NL-vdW}.

For the lattice constant, the NL-vdW MGGA SCAN+rVV10 and MGGA TM lead
to the lowest MAE (0.02~\AA) and MARE (0.5\%). Without the nonlocal dispersion term,
the MAE and MARE with SCAN only slightly increases to 0.03~\AA~and 0.6\%,
which are the same values obtained with the GGA PBEsol.
Note that other accurate GGA functionals like SG4\cite{ConstantinPRB16} or
WC\cite{WuPRB06} also lead to errors in this range (see Ref.~\onlinecite{TranJCP16}).
Interestingly, the ME and MRE for SCAN+rVV10 are basically zero, which means that
this functional does not show a particular tendency towards underestimation or
overestimation of $a_{0}$.
Most of the other modern functionals lead to values that are in the range
0.04-0.05~\AA~for the MAE and 0.8-1.0\% for the MARE.
The results obtained with the recent PBEsol+rVV10s can be considered as very accurate
and are overall only slightly deteriorated with respect to those obtained with
the GGA PBEsol without dispersion correction. As already known from
previous studies (see, e.g., Refs.~\onlinecite{KlimesPRB11,ParkCAP15}), 
the first two original NL-vdW functionals
vdW-DF and vdW-DF2 lead to very large overestimations, similarly as the worst
GGAs for solids like BLYP\cite{BeckePRA88,LeePRB88} do.\cite{TranJCP16}
This is due to the strong magnitude of the enhancement factors $F_{\text{xc}}$ of the
corresponding semilocal components (see Fig.~\ref{fig_Fxc}) which favor too much
inhomogeneities in $\rho$, and therefore too large equilibrium volumes,
in the case of solids. rVV10 also shows a clear overestimation of the
lattice constant which is nearly as large as with PBE. From Fig.~\ref{fig_Fxc},
we can see that the factor $F_{\text{xc}}$ is slightly larger in the case
of rVV10 than PBE, however the additional nonlocal term in rVV10 reduces
the overestimation in $a_{0}$.

As expected, the accuracy for the bulk modulus $B_{0}$ follows a trend that
is rather similar as for the lattice constant; if a functional is among the most
accurate for $a_{0}$, then the same conclusion holds also for $B_{0}$.

The results for the cohesive energy $E_{\text{coh}}$ show that the lowest MAE
(0.13~eV/atom) and MARE ($3.2$\%) are obtained with the NL-vdW functional
rVV10. Remarkably, these mean errors are smaller than
all those obtained with the 62 functionals tested in Ref.~\onlinecite{TranJCP16}.
However, the price to pay is to have errors for the lattice constant that are
quite large, since the MAE and MARE are more than three times larger than with SCAN+rVV10.
Actually, this is a problem that is often encountered with GGAs: a functional
that is among the most accurate ones for property A will most likely not be very
accurate for another property B that is quite different from property A
(see also Sec.~\ref{molecules}). Also very accurate are optB88-vdW and rev-vdW-DF2
with MAE of 0.13-0.14~eV/atom and MARE in the range 3.8-4.0\%.
The results obtained with SCAN+rVV10 (the best one for $a_{0}$) are
relatively fair, but a certain number of other functionals perform better.
We also note the extremely bad performance (strong
overestimation) of PBEsol+rVV10s for $E_{\text{coh}}$ with MAE and MARE of
0.45~eV/atom and 10.6\%, respectively, which makes this functional the fourth
worst after LDA, vdW-DF2, and vdW-DF. Actually, as expected and already shown in
Ref.~\onlinecite{TerentjevPRB18b} PBEsol+rVV10s significantly worsens the
cohesive energy with respect to PBEsol, while it was only slightly the case for the
lattice constant.

By considering the results in Table~\ref{table_strong} as a whole, an accurate
or satisfying description of the three properties seems to be achieved by the
following functionals: SCAN, SCAN+rVV10, vdW-DF-cx, PBE-D3(BJ), optB86b-vdW, 
PBE+rVV10L, and rev-vdW-DF2,

Figures~\ref{fig_solids_1} and \ref{fig_solids_2} show the results for the lattice constant
and cohesive energy for some of the most recent NL-vdW functionals as well as the
MGGAs SCAN and TM. A few interesting observations are the following.
The underestimation (overestimation) by PBEsol+rVV10s of the lattice constant
(cohesive energy) is particularly pronounced for the alkali and alkaline earth metals.
For some of the transition metals, namely Ni, Cu, Rh, Pd, and Ir, PBEsol+rVV10s and TM
clearly overestimate $E_{\text{coh}}$. Interestingly, for the lattice constant, PBE+rVV10L
and rev-vdW-DF2 lead to very similar results except for the heavy alkali
and alkaline earth metals. We note that in general, the difference between the
SCAN and SCAN+rVV10 results is quite small, which can be inferred from
Fig.~\ref{fig_Fxc}, where we already observed that the NL-vdW term of SCAN+rVV10 has the
smallest magnitude. Finally, we mention that SCAN+rVV10 does not perform well for
Cs (strong overestimation of $a_{0}$), but leads to rather consistent
results for the semiconductors and the transition metals.

\subsection{\label{weakly}Weakly bound solids}

\subsubsection{\label{raregases}Rare-gas solids}

\begin{table*}
\caption{\label{table_RG}Equilibrium lattice constant $a_{0}$ (in \AA) and
cohesive energy $E_{\text{coh}}$ (in meV/atom) of rare-gas solids calculated
from various functionals and compared to reference CCSD(T) results.\cite{RosciszewskiPRB00}
The units of the MRE and MARE are \%.
The results were obtained from self-consistent
calculations, except those for SCAN, SCAN+rVV10, and TM that were obtained
using the PBE orbitals/density. The functionals are separated into two groups,
those which contain a dispersion term (NL or atom-pairwise), and those which do not.
Within each group, the functionals are ordered by increasing MARE.
The errors (indicated in parenthesis)
larger than 5\% for $a_{0}$ and 30\% for $E_{\text{coh}}$ are underlined.}
\begin{ruledtabular}
\begin{tabular}{lcccccccc}
Method & Ne & Ar & Kr & Xe & ME & MAE & MRE & MARE \\
\hline
\multicolumn{9}{c}{$a_{0}$} \\
Without dispersion           \\
TM  &  4.05 (\underline{-6}) &  5.23 (0) &  5.60 (0) &  6.15 (1) & -0.05 &  0.08 &  -1.2 &   1.8 \\
SCAN  &  4.03 (\underline{-6}) &  5.31 (1) &  5.74 (2) &  6.33 (4) &  0.04 &  0.18 &   0.3 &   3.4 \\
LDA  &  3.86 (\underline{-10}) &  4.94 (\underline{-6}) &  5.33 (-5) &  5.85 (-4) & -0.31 &  0.31 &  -6.2 &   6.2 \\
PBEsol  &  4.70 (\underline{9}) &  5.88 (\underline{12}) &  6.13 (\underline{10}) &  6.48 (\underline{6}) &  0.49 &  0.49 &   9.3 &   9.3 \\
PBE &  4.60 (\underline{7}) &  5.96 (\underline{13}) &  6.42 (\underline{15}) &  7.03 (\underline{16}) &  0.69 &  0.69 &  12.7 &  12.7 \\
With dispersion         \\
optB88-vdW  &  4.26 (-1) &  5.23 (0) &  5.63 (1) &  6.15 (1) &  0.01 &  0.04 &   0.1 &   0.7 \\
optB86b-vdW  &  4.35 (1) &  5.32 (1) &  5.68 (1) &  6.18 (2) &  0.07 &  0.07 &   1.4 &   1.4 \\
rVV10  &  4.21 (-2) &  5.16 (-2) &  5.52 (-1) &  6.01 (-1) & -0.08 &  0.08 &  -1.6 &   1.6 \\
PBEsol+rVV10s  &  4.41 (3) &  5.38 (2) &  5.67 (1) &  6.08 (0) &  0.08 &  0.08 &   1.6 &   1.6 \\
C09-vdW  &  4.55 (\underline{6}) &  5.34 (2) &  5.63 (1) &  6.07 (0) &  0.09 &  0.10 &   2.0 &   2.1 \\
vdW-DF2  &  4.17 (-3) &  5.28 (1) &  5.74 (2) &  6.31 (4) &  0.07 &  0.13 &   1.0 &   2.4 \\
rev-vdW-DF2  &  4.42 (3) &  5.37 (2) &  5.74 (3) &  6.22 (2) &  0.13 &  0.13 &   2.5 &   2.5 \\
SCAN+rVV10  &  3.97 (\underline{-8}) &  5.17 (-1) &  5.56 (-1) &  6.12 (1) & -0.10 &  0.12 &  -2.3 &   2.6 \\
PBE+rVV10L  &  4.37 (2) &  5.48 (4) &  5.86 (5) &  6.33 (4) &  0.20 &  0.20 &   3.6 &   3.6 \\
PBE-D3(BJ)  &  4.46 (4) &  5.49 (5) &  5.85 (5) &  6.31 (4) &  0.22 &  0.22 &   4.1 &   4.1 \\
vdW-DF  &  4.34 (1) &  5.50 (5) &  5.95 (\underline{6}) &  6.54 (\underline{7}) &  0.28 &  0.28 &   4.9 &   4.9 \\
vdW-DF-cx  &  4.40 (2) &  5.59 (\underline{7}) &  6.05 (\underline{8}) &  6.53 (\underline{7}) &  0.34 &  0.34 &   6.1 &   6.1 \\
revPBE-D3(BJ)  &  4.80 (\underline{12}) &  5.67 (\underline{8}) &  5.96 (\underline{7}) &  6.37 (5) &  0.39 &  0.39 &   7.8 &   7.8 \\
Reference\footnotemark[1] & 4.30 & 5.25 & 5.60 & 6.09 \\
\hline
\multicolumn{9}{c}{$E_{\text{coh}}$} \\
Without dispersion           \\
TM  &     47 (\underline{80}) &     62 (-30) &     82 (\underline{-33}) &     95 (\underline{-44}) &   -30 &    41 &  -6.7 &  46.6  \\
SCAN  &     54 (\underline{107}) &     61 (-30) &     72 (\underline{-41}) &     74 (\underline{-56}) &   -36 &    50 &  -5.2 &  58.7  \\
PBE  &     19 (-26) &     23 (\underline{-73}) &     27 (\underline{-78}) &     29 (\underline{-83}) &   -77 &    77 & -64.9 &  64.9  \\
PBEsol  &     12 (\underline{-54}) &     17 (\underline{-81}) &     23 (\underline{-81}) &     32 (\underline{-81}) &   -81 &    81 & -74.4 &  74.4  \\
LDA  &     87 (\underline{234}) &    138 (\underline{57}) &    169 (\underline{39}) &    202 (19) &    48 &    48 &  87.2 &  87.2  \\
With dispersion         \\
revPBE-D3(BJ)  &     25 (-2) &     82 (-7) &    126 (3) &    192 (13) &     5 &     8 &   1.7 &   6.5  \\
rev-vdW-DF2  &     31 (19) &     82 (-7) &    111 (-9) &    148 (-13) &    -9 &    11 &  -2.4 &  12.0  \\
PBE-D3(BJ)  &     37 (\underline{42}) &     86 (-2) &    117 (-4) &    162 (-5) &    -1 &     6 &   7.6 &  13.1  \\
PBEsol+rVV10s  &     29 (10) &     57 (\underline{-35}) &     75 (\underline{-39}) &    111 (\underline{-35}) &   -34 &    35 & -24.7 &  29.8  \\
PBE+rVV10L  &     45 (\underline{72}) &     79 (-10) &    102 (-17) &    130 (-24) &   -13 &    22 &   5.4 &  30.7  \\
rVV10  &     42 (\underline{60}) &    113 (28) &    162 (\underline{33}) &    226 (\underline{33}) &    34 &    34 &  38.4 &  38.4  \\
C09-vdW  &     51 (\underline{98}) &    118 (\underline{34}) &    156 (28) &    212 (25) &    33 &    33 &  46.1 &  46.1  \\
vdW-DF2  &     58 (\underline{122}) &    124 (\underline{41}) &    154 (27) &    190 (11) &    30 &    30 &  50.0 &  50.0  \\
optB88-vdW  &     50 (\underline{93}) &    138 (\underline{57}) &    180 (\underline{47}) &    234 (\underline{37}) &    49 &    49 &  58.7 &  58.7  \\
SCAN+rVV10  &     79 (\underline{204}) &    111 (26) &    137 (12) &    159 (-7) &    20 &    26 &  58.9 &  62.3  \\
optB86b-vdW  &     61 (\underline{134}) &    137 (\underline{56}) &    174 (\underline{42}) &    224 (\underline{32}) &    47 &    47 &  65.9 &  65.9  \\
vdW-DF-cx  &     79 (\underline{205}) &    137 (\underline{56}) &    160 (\underline{31}) &    191 (12) &    40 &    40 &  76.0 &  76.0  \\
vdW-DF &     92 (\underline{253}) &    156 (\underline{77}) &    181 (\underline{49}) &    212 (25) &    59 &    59 & 100.9 & 100.9  \\
Reference\footnotemark[1] & 26 & 88 & 122 & 170 \\
\end{tabular}
\end{ruledtabular}
\footnotetext[1]{The set of CCSD(T) results from Ref.~\onlinecite{RosciszewskiPRB00}
that include the two-, three-, and four-body contributions, but not the effect due to
the zero-point vibration.}
\end{table*}

Turning to weakly bound systems, Table~\ref{table_RG} shows the results for
the rare-gas solids Ne, Ar, Kr, and Xe which crystallize in the face-centered
cubic structure and have been used in many previous works
\cite{OrtmannPRB06,TranPRB07,HarlPRB08,HaasPRB09a,YousafJCTC10,BuckoJPCA10,AlSaidiJCTC12,BuckoPRB13,TranJCP13,MoellmannJPCC14,CallsenPRB15,TranJCP16,PengPRX16,TerentjevC18,TranJCP18,TerentjevPRB18b}
for testing functionals since they represent the prototypical van der Waals
systems bound by dispersion forces. As criteria to decide what is an
(unacceptably) large error with respect to the very accurate
CCSD(T) (coupled cluster with singlet, doublet,
and perturbative triplet) values,\cite{RosciszewskiPRB00} we chose 5\% and 30\% for the relative error
on the lattice constant and cohesive energy, respectively.

The results show that only one method, the NL-vdW functional rev-vdW-DF2, leads
to no such large errors as defined by our criteria (see also Ref.~\onlinecite{CallsenPRB15}).
The largest error is 3\% for $a_{0}$ (Ne and Kr) and 19\% for $E_{\text{coh}}$ (Ne).
Another functional which also performs rather well and shows only one large
error is the atom-pairwise PBE-D3(BJ), which leads to an overestimation of 42\% for
$E_{\text{coh}}$ of Ne, but below 5\% for the others. The other
atom-pairwise functional, revPBE-D3(BJ), leads to very accurate cohesive energy
for the four rare gases, but is overall one of the most inaccurate methods
for the lattice constant. Somehow satisfying overall, PBEsol+rVV10s leads
to errors in the range 35-40\% for the cohesive energy of Ar, Kr, and Xe, but
only 10\% for Ne.

All other functionals lead to at least one very large error above 50\% for
$E_{\text{coh}}$, including SCAN+rVV10 that performs very badly for Ne for
both $a_{0}$ and $E_{\text{coh}}$, as already noticed in
Ref.~\onlinecite{PengPRX16}. Thus, except rev-vdW-DF2, PBE-D3(BJ), and
PBEsol+rVV10s, none of the other functionals can be considered as satisfying
for rare-gas solids. However, note that other non-hybrid DFT+D3 or DFT+D3(BJ)
methods, e.g. PBEsol-D3(BJ)\cite{GoerigkPCCP11} or BLYP-D3,\cite{GrimmeJCP10}
can also be reliable for the rare gases, as shown in our previous
work.\cite{TranJCP16} We can also see that the most accurate methods for $a_{0}$,
optB88-vdW and optB86b-vdW, which lead to errors not larger than $\sim1\%$
are extremely inaccurate for $E_{\text{coh}}$ in all cases.

\subsubsection{\label{layered}Layered solids}

\begin{table*}
\caption{\label{table_LC}Equilibrium lattice constants (intralayer $a_{0}$ and
interlayer $c_{0}$, in \AA) and interlayer binding energy [$E_{b}$,
in meV/\AA$^{2}$, i.e., meV per surface area $A=a_{0}^{2}\cos(\pi/6)$ in the bulk]
of layered solids. The units of the MRE and MARE are \%.
The reference results are from experiment for $a_{0}$ and $c_{0}$
(with zero-point vibration effect removed) and
from RPA calculations for $E_{b}$ (see Ref.~\onlinecite{BjorkmanJCP14}).
The results were obtained from self-consistent
calculations, except those for SCAN, SCAN+rVV10, and TM that were obtained
using the PBE orbitals/density. The functionals are separated into two groups,
those which contain a dispersion term (NL or atom-pairwise), and those which do not.
Within each group, the functionals are ordered by increasing MARE.
The errors (indicated in parenthesis)
larger than 1\% for $a_{0}$, 3\% for $c_{0}$, and 30\% for $E_{b}$ are underlined.}
{\scriptsize
\begin{ruledtabular}
\begin{tabular}{lcccccccccccccc}
Method        & Graphite & h-BN  & TiS$_{2}$ & TiSe$_{2}$ & MoS$_{2}$ & MoSe$_{2}$ & MoTe$_{2}$ & HfTe$_{2}$ & WS$_{2}$ & WSe$_{2}$ & ME & MAE & MRE & MARE \\
\hline
\multicolumn{15}{c}{$a_{0}$} \\
Without dispersion \\
SCAN          &   2.45 (0) &   2.50 (0) &   3.42 (0) &   3.54 (0) &   3.18 (1) &   3.30 (0) &   3.53 (0) &   3.97 (0) &   3.17 (1) &   3.30 (1) &   0.01 &   0.01 &    0.2 &    0.4 \\
TM            &   2.46 (0) &   2.51 (0) &   3.38 (-1) &   3.50 (-1) &   3.15 (0) &   3.27 (-1) &   3.49 (-1) &   3.90 (-1) &   3.15 (0) &   3.27 (0) &  -0.02 &   0.02 &   -0.5 &    0.6 \\
PBE           &   2.47 (1) &   2.51 (0) &   3.42 (0) &   3.54 (0) &   3.19 (1) &   3.32 (1) &   3.56 (1) &   3.98 (1) &   3.19 (1) &   3.32 (1) &   0.02 &   0.02 &    0.7 &    0.7 \\
PBEsol        &   2.46 (0) &   2.50 (0) &   3.35 (\underline{-2}) &   3.48 (\underline{-2}) &   3.14 (-1) &   3.27 (-1) &   3.50 (-1) &   3.88 (\underline{-2}) &   3.15 (0) &   3.27 (0) &  -0.03 &   0.03 &   -0.8 &    0.8 \\
LDA           &   2.45 (0) &   2.49 (-1) &   3.31 (\underline{-3}) &   3.43 (\underline{-3}) &   3.12 (-1) &   3.25 (-1) &   3.47 (-1) &   3.82 (\underline{-3}) &   3.13 (-1) &   3.25 (-1) &  -0.06 &   0.06 &   -1.6 &    1.6 \\
With dispersion \\
SCAN+rVV10    &   2.45 (0) &   2.50 (0) &   3.41 (0) &   3.54 (0) &   3.17 (0) &   3.29 (0) &   3.52 (0) &   3.95 (0) &   3.16 (0) &   3.29 (0) &   0.00 &   0.01 &    0.0 &    0.2 \\
PBE+rVV10L    &   2.46 (0) &   2.51 (0) &   3.39 (-1) &   3.52 (0) &   3.17 (0) &   3.30 (0) &   3.53 (0) &   3.94 (0) &   3.17 (1) &   3.30 (1) &   0.00 &   0.01 &    0.1 &    0.4 \\
optB86b-vdW   &   2.46 (0) &   2.51 (0) &   3.38 (-1) &   3.52 (0) &   3.17 (0) &   3.30 (0) &   3.53 (0) &   3.94 (0) &   3.17 (1) &   3.30 (1) &   0.00 &   0.01 &    0.0 &    0.4 \\
rev-vdW-DF2   &   2.46 (0) &   2.51 (0) &   3.39 (-1) &   3.52 (0) &   3.17 (0) &   3.31 (1) &   3.54 (1) &   3.95 (0) &   3.17 (1) &   3.30 (1) &   0.00 &   0.01 &    0.2 &    0.4 \\
vdW-DF-cx     &   2.46 (0) &   2.51 (0) &   3.36 (-1) &   3.49 (-1) &   3.15 (0) &   3.28 (0) &   3.51 (0) &   3.90 (-1) &   3.15 (0) &   3.28 (0) &  -0.02 &   0.02 &   -0.5 &    0.5 \\
PBE-D3(BJ)    &   2.46 (0) &   2.51 (0) &   3.36 (-1) &   3.49 (-1) &   3.16 (0) &   3.28 (0) &   3.50 (-1) &   3.90 (-1) &   3.19 (1) &   3.29 (0) &  -0.01 &   0.02 &   -0.3 &    0.7 \\
optB88-vdW    &   2.46 (0) &   2.51 (0) &   3.41 (0) &   3.55 (0) &   3.19 (1) &   3.32 (1) &   3.58 (\underline{2}) &   3.99 (1) &   3.19 (1) &   3.32 (1) &   0.02 &   0.02 &    0.7 &    0.7 \\
C09-vdW       &   2.46 (0) &   2.51 (0) &   3.35 (\underline{-2}) &   3.48 (\underline{-2}) &   3.14 (-1) &   3.27 (-1) &   3.49 (-1) &   3.88 (\underline{-2}) &   3.15 (0) &   3.27 (0) &  -0.03 &   0.03 &   -0.8 &    0.8 \\
revPBE-D3(BJ) &   2.47 (1) &   2.51 (0) &   3.34 (\underline{-2}) &   3.46 (\underline{-2}) &   3.13 (-1) &   3.25 (-1) &   3.47 (-1) &   3.85 (\underline{-3}) &   3.14 (0) &   3.26 (-1) &  -0.04 &   0.04 &   -1.1 &    1.2 \\
PBEsol+rVV10s &   2.46 (0) &   2.50 (0) &   3.33 (\underline{-2}) &   3.46 (\underline{-2}) &   3.13 (-1) &   3.25 (-1) &   3.47 (-1) &   3.83 (\underline{-3}) &   3.13 (-1) &   3.26 (-1) &  -0.05 &   0.05 &   -1.3 &    1.3 \\
rVV10         &   2.47 (1) &   2.52 (0) &   3.44 (1) &   3.58 (1) &   3.22 (\underline{2}) &   3.36 (\underline{2}) &   3.60 (\underline{2}) &   4.02 (\underline{2}) &   3.22 (\underline{2}) &   3.36 (\underline{2}) &   0.05 &   0.05 &    1.6 &    1.6 \\
vdW-DF        &   2.48 (1) &   2.52 (0) &   3.48 (\underline{2}) &   3.62 (\underline{2}) &   3.24 (\underline{2}) &   3.38 (\underline{3}) &   3.64 (\underline{3}) &   4.08 (\underline{3}) &   3.24 (\underline{3}) &   3.38 (\underline{3}) &   0.08 &   0.08 &    2.3 &    2.3 \\
vdW-DF2       &   2.48 (1) &   2.52 (0) &   3.52 (\underline{3}) &   3.68 (\underline{4}) &   3.29 (\underline{4}) &   3.45 (\underline{5}) &   3.72 (\underline{6}) &   4.16 (\underline{5}) &   3.29 (\underline{4}) &   3.44 (\underline{5}) &   0.13 &   0.13 &    3.8 &    3.8 \\
Reference     &   2.46 &   2.51 &   3.41 &   3.54 &   3.16 &   3.29 &   3.52 &   3.96 &   3.15 &   3.28 \\
\hline
\multicolumn{15}{c}{$c_{0}$} \\
Without dispersion \\
TM            &  6.63 (0) &  6.51 (-2) &  5.76 (1) &  6.12 (2) &  12.5 (2) &  13.2 (2) &  14.2 (2) &  6.75 (2) &  12.6 (2) &  13.2 (2) &    0.2 &    0.2 &    1.4 &    1.7 \\
LDA           &  6.63 (0) &  6.49 (-2) &  5.45 (\underline{-4}) &  5.80 (-3) &  12.1 (-2) &  12.8 (-1) &  13.8 (-1) &  6.50 (-2) &  12.2 (-1) &  12.8 (-1) &   -0.2 &    0.2 &   -1.8 &    1.8 \\
PBEsol        &  7.26 (\underline{9}) &  7.06 (\underline{6}) &  5.65 (-1) &  5.92 (-1) &  12.6 (3) &  13.1 (2) &  14.0 (0) &  6.60 (-1) &  12.7 (3) &  13.2 (2) &    0.2 &    0.2 &    2.3 &    2.9 \\
SCAN          &  6.95 (\underline{5}) &  6.82 (3) &  5.93 (\underline{4}) &  6.32 (\underline{5}) &  12.9 (\underline{5}) &  13.6 (\underline{5}) &  14.7 (\underline{5}) &  6.97 (\underline{5}) &  12.9 (\underline{5}) &  13.6 (\underline{5}) &    0.5 &    0.5 &    4.8 &    4.8 \\
PBE           &  8.84 (\underline{33}) &  8.69 (\underline{31}) &  6.61 (\underline{16}) &  6.70 (\underline{12}) &  14.8 (\underline{21}) &  15.1 (\underline{17}) &  15.3 (\underline{10}) &  7.21 (\underline{9}) &  14.9 (\underline{21}) &  15.2 (\underline{18}) &    1.7 &    1.7 &   18.7 &   18.7 \\
With dispersion \\
rev-vdW-DF2   &  6.64 (0) &  6.57 (-1) &  5.68 (0) &  6.00 (0) &  12.4 (1) &  13.1 (1) &  14.1 (1) &  6.71 (1) &  12.4 (1) &  13.2 (2) &    0.1 &    0.1 &    0.6 &    0.9 \\
optB86b-vdW   &  6.63 (0) &  6.53 (-2) &  5.69 (0) &  6.00 (0) &  12.4 (1) &  13.1 (1) &  14.1 (1) &  6.70 (1) &  12.5 (1) &  13.2 (2) &    0.1 &    0.1 &    0.5 &    0.9 \\
vdW-DF-cx     &  6.56 (-1) &  6.45 (-3) &  5.61 (-2) &  5.93 (-1) &  12.3 (0) &  12.9 (0) &  13.9 (0) &  6.60 (-1) &  12.4 (1) &  13.0 (1) &   -0.0 &    0.1 &   -0.6 &    0.9 \\
PBEsol+rVV10s &  6.70 (1) &  6.60 (-1) &  5.54 (-3) &  5.87 (-2) &  12.2 (-1) &  12.9 (0) &  13.8 (-1) &  6.57 (-1) &  12.3 (0) &  13.0 (0) &   -0.1 &    0.1 &   -0.7 &    0.9 \\
rVV10         &  6.71 (1) &  6.62 (0) &  5.70 (0) &  6.05 (1) &  12.4 (1) &  13.1 (2) &  14.2 (2) &  6.75 (2) &  12.5 (1) &  13.2 (2) &    0.1 &    0.1 &    1.1 &    1.1 \\
C09-vdW       &  6.46 (-3) &  6.35 (\underline{-4}) &  5.55 (-3) &  5.89 (-2) &  12.2 (-1) &  12.8 (0) &  13.9 (-1) &  6.57 (-1) &  12.2 (0) &  12.9 (0) &   -0.1 &    0.1 &   -1.5 &    1.5 \\
PBE-D3(BJ)    &  6.79 (2) &  6.68 (1) &  5.56 (-2) &  5.59 (\underline{-7}) &  12.2 (-1) &  12.8 (0) &  13.8 (-1) &  6.58 (-1) &  12.2 (0) &  12.9 (0) &   -0.1 &    0.1 &   -1.0 &    1.6 \\
optB88-vdW    &  6.69 (1) &  6.60 (-1) &  5.75 (1) &  6.14 (2) &  12.5 (2) &  13.2 (2) &  14.3 (2) &  6.80 (2) &  12.6 (2) &  13.3 (2) &    0.2 &    0.2 &    1.7 &    1.8 \\
SCAN+rVV10    &  6.68 (1) &  6.59 (-1) &  5.75 (1) &  6.22 (\underline{4}) &  12.5 (1) &  13.2 (2) &  14.3 (2) &  6.82 (3) &  12.6 (2) &  13.2 (2) &    0.2 &    0.2 &    1.8 &    1.9 \\
PBE+rVV10L    &  6.98 (\underline{5}) &  6.88 (\underline{4}) &  5.78 (1) &  6.04 (1) &  12.6 (2) &  13.2 (2) &  14.1 (1) &  6.71 (1) &  12.7 (3) &  13.2 (2) &    0.2 &    0.2 &    2.3 &    2.3 \\
revPBE-D3(BJ) &  6.45 (-3) &  6.34 (\underline{-4}) &  5.40 (\underline{-5}) &  5.76 (\underline{-4}) &  11.8 (\underline{-4}) &  12.5 (-3) &  13.5 (-3) &  6.50 (-2) &  11.8 (\underline{-4}) &  12.5 (-3) &   -0.3 &    0.3 &   -3.6 &    3.6 \\
vdW-DF2       &  7.06 (\underline{6}) &  6.99 (\underline{5}) &  5.96 (\underline{5}) &  6.36 (\underline{6}) &  12.9 (\underline{5}) &  13.7 (\underline{6}) &  14.9 (\underline{7}) &  7.07 (\underline{6}) &  13.0 (\underline{5}) &  13.7 (\underline{6}) &    0.6 &    0.6 &    5.8 &    5.8 \\
vdW-DF        &  7.19 (\underline{8}) &  7.12 (\underline{7}) &  6.11 (\underline{7}) &  6.50 (\underline{8}) &  13.2 (\underline{7}) &  13.9 (\underline{7}) &  15.0 (\underline{8}) &  7.18 (\underline{8}) &  13.2 (\underline{7}) &  13.9 (\underline{8}) &    0.7 &    0.7 &    7.7 &    7.7 \\
Reference     &  6.63 &  6.63 &  5.70 &  6.00 & 12.28 & 12.91 & 13.96 &  6.64 & 12.31 & 12.95 \\
\hline
\multicolumn{15}{c}{$E_{b}$} \\
Without dispersion \\
LDA           &    10 (\underline{-48}) &    10 (-30) &    20 (7) &    21 (24) &    13 (\underline{-35}) &    14 (-29) &    15 (-26) &    19 (3) &    13 (\underline{-37}) &    13 (\underline{-32}) &     -4 &      5 &  -20.4 &   27.4 \\
TM            &    11 (\underline{-38}) &    12 (-19) &    13 (\underline{-31}) &    14 (-19) &    10 (\underline{-50}) &    11 (\underline{-42}) &    13 (\underline{-36}) &    13 (-28) &    10 (\underline{-50}) &    11 (\underline{-44}) &     -7 &      7 &  -35.7 &   35.7 \\
SCAN          &     7 (\underline{-59}) &     8 (\underline{-45}) &     6 (\underline{-68}) &     6 (\underline{-64}) &     6 (\underline{-73}) &     5 (\underline{-72}) &     7 (\underline{-65}) &     7 (\underline{-60}) &     6 (\underline{-72}) &     5 (\underline{-73}) &    -12 &     12 &  -65.2 &   65.2 \\
PBEsol        &     2 (\underline{-92}) &     2 (\underline{-86}) &     7 (\underline{-62}) &    10 (\underline{-44}) &     3 (\underline{-84}) &     5 (\underline{-75}) &     8 (\underline{-62}) &    10 (\underline{-45}) &     3 (\underline{-86}) &     4 (\underline{-77}) &    -13 &     13 &  -71.4 &   71.4 \\
PBE           &     1 (\underline{-97}) &     1 (\underline{-96}) &     1 (\underline{-93}) &     2 (\underline{-90}) &     1 (\underline{-97}) &     1 (\underline{-97}) &     1 (\underline{-94}) &     2 (\underline{-90}) &     1 (\underline{-97}) &     1 (\underline{-97}) &    -18 &     18 &  -94.8 &   94.8 \\
With dispersion \\
SCAN+rVV10    &    20 (7) &    19 (\underline{34}) &    18 (-3) &    18 (3) &    20 (-3) &    19 (-1) &    21 (2) &    19 (0) &    21 (4) &    20 (-1) &      1 &      1 &    4.3 &    5.8 \\
PBE+rVV10L    &    15 (-19) &    14 (-4) &    19 (2) &    20 (18) &    19 (-6) &    20 (2) &    22 (4) &    20 (6) &    19 (-5) &    20 (1) &     -0 &      1 &   -0.1 &    6.7 \\
vdW-DF2       &    20 (8) &    19 (29) &    19 (1) &    18 (2) &    19 (-6) &    18 (-9) &    16 (-21) &    15 (-19) &    19 (-5) &    18 (-10) &     -1 &      2 &   -3.0 &   10.9 \\
vdW-DF        &    20 (12) &    19 (\underline{35}) &    19 (0) &    18 (2) &    19 (-7) &    18 (-10) &    16 (-21) &    15 (-18) &    19 (-5) &    18 (-11) &     -1 &      2 &   -2.3 &   12.0 \\
PBEsol+rVV10s &    12 (\underline{-32}) &    12 (-20) &    20 (7) &    21 (21) &    17 (-15) &    17 (-11) &    21 (3) &    22 (19) &    17 (-14) &    17 (-12) &     -1 &      3 &   -5.4 &   15.3 \\
rev-vdW-DF2   &    23 (23) &    21 (\underline{47}) &    25 (30) &    24 (\underline{40}) &    23 (14) &    22 (15) &    23 (9) &    22 (16) &    23 (14) &    22 (12) &      4 &      4 &   22.0 &   22.0 \\
PBE-D3(BJ)    &    17 (-9) &    16 (9) &    27 (\underline{45}) &    30 (\underline{72}) &    24 (17) &    26 (\underline{34}) &    30 (\underline{44}) &    27 (\underline{46}) &    26 (28) &    28 (\underline{38}) &      6 &      7 &   32.5 &   34.2 \\
vdW-DF-cx     &    25 (\underline{36}) &    24 (\underline{67}) &    27 (\underline{43}) &    27 (\underline{59}) &    25 (21) &    25 (26) &    26 (25) &    25 (\underline{35}) &    24 (21) &    24 (23) &      6 &      6 &   35.6 &   35.6 \\
optB88-vdW    &    27 (\underline{47}) &    26 (\underline{80}) &    27 (\underline{45}) &    26 (\underline{52}) &    26 (28) &    25 (29) &    24 (16) &    23 (23) &    26 (29) &    25 (27) &      7 &      7 &   37.5 &   37.5 \\
optB86b-vdW   &    27 (\underline{47}) &    26 (\underline{80}) &    28 (\underline{48}) &    28 (\underline{60}) &    26 (29) &    26 (\underline{31}) &    26 (24) &    25 (\underline{33}) &    26 (30) &    26 (28) &      7 &      7 &   41.0 &   41.0 \\
rVV10         &    26 (\underline{44}) &    25 (\underline{72}) &    28 (\underline{48}) &    29 (\underline{65}) &    29 (\underline{42}) &    29 (\underline{50}) &    29 (\underline{40}) &    26 (\underline{40}) &    29 (\underline{44}) &    29 (\underline{48}) &      9 &      9 &   49.2 &   49.2 \\
C09-vdW       &    29 (\underline{59}) &    28 (\underline{96}) &    32 (\underline{72}) &    33 (\underline{88}) &    30 (\underline{44}) &    29 (\underline{49}) &    30 (\underline{44}) &    30 (\underline{59}) &    29 (\underline{44}) &    29 (\underline{45}) &     11 &     11 &   59.9 &   59.9 \\
revPBE-D3(BJ) &    26 (\underline{41}) &    25 (\underline{71}) &    48 (\underline{153}) &    51 (\underline{196}) &    45 (\underline{118}) &    49 (\underline{152}) &    55 (\underline{163}) &    46 (\underline{147}) &    50 (\underline{147}) &    53 (\underline{167}) &     26 &     26 &  135.5 &  135.5 \\
Reference     &  18.3 &  14.4 &  18.8 &  17.3 &  20.5 &  19.6 &  20.8 &  18.6 &  20.2 &  19.9 \\
\end{tabular}
\end{ruledtabular}
}
\end{table*}

\begin{figure}
\includegraphics[scale=0.65]{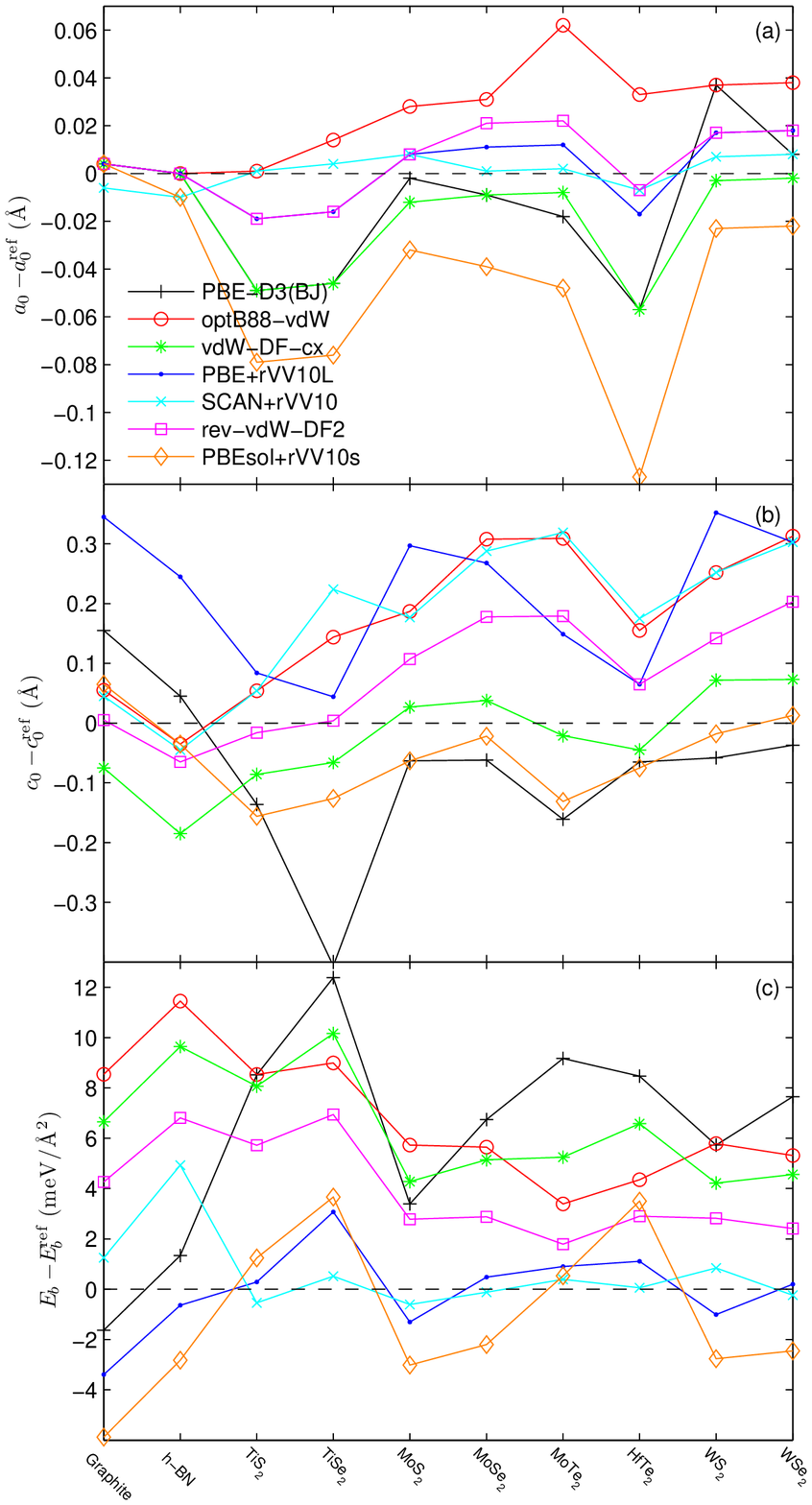}
\caption{\label{fig_LC}Error for the intralayer and interlayer lattice constants [in (a)
and (b), respectively] and binding energy [(c)] shown for selected functionals.}
\end{figure}

The hexagonal layered solids constitute another set of prototypical systems
bound by weak interactions that is often used for assessing functionals.
\cite{HasegawaPRB04,OrtmannPRB06,HasegawaPRB07,TranPRB07,HaasPRB09a,BjorkmanPRL12,BjorkmanJPCM12,BjorkmanPRB12,BjorkmanJCP14,GrazianoJPCM12,BerlandPRB14,HamadaPRB14,BuckoPRB13,RegoJPCM15,TranJCP16,PengPRX16,LebedevaCMS17,PengPRB17,TerentjevC18,TawfikPRM18,MejiaRodriguezPRB18,MosyaginPRB18,TerentjevPRB18b}
These systems consist of hexagonal layers that are bound by weak interactions,
while the atoms within a layer are strongly bound. The results for the
intralayer and interlayer lattice constants $a_{0}$ and $c_{0}$ as well as the
interlayer binding energy $E_{b}$ are shown in Table~\ref{table_LC}
and compared to results obtained from experiment for $a_{0}$ and $c_{0}$ or
the random-phase approximation (RPA) for $E_{b}$.\cite{BjorkmanJCP14}

For selected functionals, the results are also compared graphically in Fig.~\ref{fig_LC}.
We mention that in the calculation on the monolayer to get $E_{b}$,
the intralayer lattice constant $a$ was also optimized (results not shown).
However, we observed that in the vast majority of cases choosing
either $a=a_{0}^{\text{monolayer}}$ or $a=a_{0}^{\text{bulk}}$ for the
monolayer has a very small influence, a few tenths of meV/\AA$^{2}$, on $E_{b}$.

The trends observed among the functionals for $a_{0}$ are, as expected,
similar to those for the strongly bound solids discussed above. In brief,
the largest underestimations (up to a few percents) are due to LDA, revPBE-D3(BJ),
PBEsol+rVV10s, and C09-vdW, while vdW-DF2 leads to very large overestimations
(up to 6\%).
vdW-DF and rVV10 also show a clear tendency towards overestimation of $a_{0}$
(see also Ref.~\onlinecite{BjorkmanPRB12}).
All other functionals perform clearly better and, as shown in Fig.~\ref{fig_LC},
the most accurate one is SCAN+rVV10 which leads to errors below
0.01~\AA~(below 0.5\%) for all systems.
For the interlayer lattice constant $c_{0}$, the functionals, beside PBE
which barely binds the layers, that can be identified
as more inaccurate than the others are vdW-DF, vdW-DF2, and SCAN, which clearly
overestimate $c_{0}$, as well as revPBE-D3(BJ) which does the opposite.
For these functionals, the error is at least 4\% for a certain number of
solids. The other functionals lead to errors which are at most 3\% for all or
most solids.

Thus, overall vdW-DF, vdW-DF2, and revPBE-D3(BJ) perform very poorly for
both $a_{0}$ and $c_{0}$. rVV10 is also among the inaccurate methods for $a_{0}$,
but performs quite well for $c_{0}$. As well-known, LDA systematically
underestimates the
lattice constant, but does it moderately for $c_{0}$ since the errors are quite
small. Most other dispersion-corrected functionals can be considered as satisfying
for both lattice constants. Note that in the work of
Bj\"{o}rkman,\cite{BjorkmanJCP14} optB86b-vdW, vdW-DF-cx, and rev-vdW-DF2
were already shown to be accurate for the lattice constants.

As observed above for the rare-gas solids, the relative errors for the
interlayer binding energy $E_{b}$ are much larger than for the lattice
constants. By considering 30\% of relative error as the largest acceptable
value for $E_{b}$, the results in Table~\ref{table_LC} show that the
functionals which have a reasonable accuracy for all solids except possibly
one are SCAN+rVV10, PBE+rVV10L, vdW-DF2, vdW-DF, and PBEsol+rVV10s.
rev-vdW-DF2 can also be considered as accurate since for only two solids (h-BN and
TiSe$_{2}$) the relative error is above 30\%.
Note that the parameter $b=10$ in the PBE+rVV10L kernel was
tuned in order to reproduce at best the RPA results for $E_{b}$, therefore
its good performance is hardly surprising.
The worst functionals are PBE and revPBE-D3(BJ); PBE gives nearly no binding,
while revPBE-D3(BJ) overestimates $E_{b}$ by more than
100\% in most cases. Such huge overestimations obtained with revPBE-D3(BJ) have
already been observed in the case of adsorption of benzene on transition-metal
surfaces.\cite{ReckienBJOC14} Other very inaccurate functionals are PBEsol, SCAN,
C09-vdW, and rVV10 as already shown in Refs.~\onlinecite{BjorkmanJPCM12,BjorkmanJCP14}
for the latter two.

By considering all results for the layered solids, the best functionals are
PBE+rVV10L, SCAN+rVV10, and rev-vdW-DF2 since they belong to the accurate
methods for $a_{0}$, $c_{0}$, and $E_{b}$ at the same time.
The results with optB88-vdW and vdW-DF-cx can also be considered
as fair. Note the curious performances of vdW-DF and vdW-DF2:
very accurate for the interlayer binding energy,
but the worst for both lattice constants.\cite{BjorkmanJPCM12}
For these two functionals, the large contribution of the NL-vdW term to $E_{b}$ 
(see Fig.~\ref{fig_kernel}) leads to an appropriate binding energy, but the
corresponding slope is not steep enough to shorten the lattice constant
sufficiently.

We also mention that among a dozen of dispersion-corrected functionals
of various families, Tawfik \textit{et al}.\cite{TawfikPRM18} concluded
that SCAN+rVV10 is overall the most accurate one for a set of
twelve layered solids (quite similar to our test set).
However, PBE+rVV10L and rev-vdW-DF2 were not considered in their work.

Since results on the layered solids were already available in the literature
for many of the functionals, it may be interesting to compare some of them with
ours. Peng \textit{et al}.\cite{PengPRX16,PengPRB17} reported
results for rev-vdW-DF2, PBE+rVV10L, and SCAN+rVV10 that were obtained
with \textsc{VASP}. For $a_{0}$, their results are in good agreement with ours
since they differ by at most 0.01~\AA.
The agreement for $c_{0}$ is relatively good for rev-vdW-DF2 and
PBE+rVV10L since the difference is typically below 0.05~\AA. A difference
of 0.05~\AA~should be considered as acceptable for such large
lattice constants determined by weak interactions.
However, with SCAN+rVV10 the disagreement for $c_{0}$ is larger (in the range
0.1-0.2~\AA), which should be due to self-consistent effects
(see discussion in Ref.~\onlinecite{TranJCP16}).
Our calculations involving MGGA functionals were done using PBE(+rVV10) for
the potential, while those from \textsc{VASP} calculations were probably done
self-consistently. The agreement for $E_{b}$ is good for the three
functionals since the discrepancies are below 1 meV/\AA$^{2}$~in all cases.
Considering now the results from Bj\"{o}rkman\cite{BjorkmanJCP14} for
six functionals (e.g., rev-vdW-DF2 or optB88-vdW) obtained with \textsc{VASP},
the results are also in fair agreement with ours for the lattice constants.
However, sizable discrepancies are observed for $E_{b}$, since his results
are consistently smaller by 2-3 meV/\AA$^{2}$ compared to our results
which agree quite well with those from Peng
\textit{et al}.\cite{PengPRX16,PengPRB17} and
Berland and Hyldgaard.\cite{BerlandPRB14}

\subsubsection{\label{molecular}Molecular solids}

\begin{table*}
\caption{\label{table_MS}Equilibrium lattice constant $a_{0}$ (in \AA) and
lattice energy $E_{\text{latt}}$ (in eV/molecule) of molecular
solids calculated from various functionals and compared to experimental results.
The results were obtained from self-consistent
calculations, except those for SCAN, SCAN+rVV10, and TM that were obtained
using the PBE orbitals/density. The functionals are separated into two groups,
those which contain a dispersion term (NL or atom-pairwise), and those which do not.
Within each group, the functionals are ordered by increasing MARE of $E_{\text{latt}}$.
The errors (indicated in parenthesis)
larger than 3\% for $a_{0}$ and 15\% for $E_{\text{latt}}$ are underlined.}
\begin{ruledtabular}
\begin{tabular}{lcccccc}
\multicolumn{1}{l}{} &
\multicolumn{2}{c}{NH$_{3}$} &
\multicolumn{2}{c}{CO$_{2}$} &
\multicolumn{2}{c}{C$_{6}$H$_{12}$N$_{4}$} \\
\cline{2-3}\cline{4-5}\cline{6-7}
\multicolumn{1}{l}{Method} &
\multicolumn{1}{c}{$a_{0}$} &
\multicolumn{1}{c}{$E_{\text{latt}}$} &
\multicolumn{1}{c}{$a_{0}$} &
\multicolumn{1}{c}{$E_{\text{latt}}$} &
\multicolumn{1}{c}{$a_{0}$} &
\multicolumn{1}{c}{$E_{\text{latt}}$} \\
\hline
Without dispersion \\
TM            &   4.98 (-1) &   0.39 (1) &   5.49 (-1) &   0.25 (-15) &   6.86 (-1) &   0.68 (\underline{-24}) \\
SCAN          &   4.98 (-1) &   0.38 (-3) &   5.53 (-1) &   0.27 (-8) &   6.99 (1) &   0.55 (\underline{-38}) \\
LDA           &   4.73 (\underline{-6}) &   0.67 (\underline{73}) &   5.28 (\underline{-5}) &   0.36 (\underline{22}) &   6.72 (-3) &   0.90 (1) \\
PBEsol        &   4.96 (-2) &   0.37 (-4) &   5.82 (\underline{5}) &   0.11 (\underline{-63}) &   7.09 (3) &   0.32 (\underline{-64}) \\
PBE           &   5.17 (2) &   0.29 (\underline{-25}) &   6.07 (\underline{9}) &   0.10 (\underline{-66}) &   7.43 (\underline{8}) &   0.24 (\underline{-74}) \\
With dispersion \\
rev-vdW-DF2   &   5.01 (-1) &   0.41 (6) &   5.61 (1) &   0.27 (-9) &   6.92 (0) &   0.91 (2) \\
revPBE-D3(BJ) &   5.06 (0) &   0.38 (-2) &   5.87 (\underline{6}) &   0.27 (-8) &   6.96 (1) &   0.82 (-8) \\
vdW-DF-cx     &   5.07 (0) &   0.41 (7) &   5.85 (\underline{5}) &   0.32 (10) &   7.01 (1) &   1.02 (14) \\
vdW-DF2       &   5.15 (2) &   0.41 (7) &   5.61 (1) &   0.34 (\underline{16}) &   7.02 (2) &   0.97 (9) \\
SCAN+rVV10    &   4.89 (-3) &   0.44 (15) &   5.44 (-2) &   0.35 (\underline{19}) &   6.84 (-1) &   0.89 (0) \\
PBE-D3(BJ)    &   5.02 (-1) &   0.43 (13) &   5.74 (3) &   0.26 (-12) &   6.99 (1) &   0.81 (-10) \\
vdW-DF        &   5.28 (\underline{5}) &   0.38 (0) &   5.81 (\underline{4}) &   0.37 (\underline{25}) &   7.16 (\underline{4}) &   1.02 (15) \\
PBE+rVV10L    &   5.07 (0) &   0.40 (4) &   5.74 (3) &   0.23 (\underline{-23}) &   7.03 (2) &   0.76 (-15) \\
rVV10         &   4.96 (-2) &   0.47 (\underline{23}) &   5.50 (-1) &   0.32 (9) &   6.86 (-1) &   1.14 (\underline{28}) \\
C09-vdW       &   4.91 (-3) &   0.47 (\underline{23}) &   5.53 (-1) &   0.33 (13) &   6.83 (-1) &   1.18 (\underline{32}) \\
optB86b-vdW   &   4.98 (-1) &   0.46 (\underline{20}) &   5.58 (0) &   0.35 (\underline{20}) &   6.91 (0) &   1.17 (\underline{31}) \\
PBEsol+rVV10s &   4.87 (\underline{-4}) &   0.47 (\underline{21}) &   5.59 (1) &   0.20 (\underline{-33}) &   6.89 (0) &   0.70 (\underline{-22}) \\
optB88-vdW    &   4.98 (-1) &   0.47 (\underline{21}) &   5.53 (-1) &   0.37 (\underline{26}) &   6.90 (0) &   1.21 (\underline{35}) \\
Reference     &   5.05\footnotemark[1] ($T=2$~K) &   0.39\footnotemark[1] &   5.56\footnotemark[2] ($T\sim5$~K) &   0.29\footnotemark[1] &   6.91\footnotemark[3] ($T=34$~K) &   0.89\footnotemark[1] \\
\end{tabular}
\end{ruledtabular}
\footnotetext[1]{Reference~\onlinecite{ReillyJCP13}. The values for $E_{\text{latt}}$
are corrected for the thermal and zero-point effects.}
\footnotetext[2]{Reference~\onlinecite{HeitCS16}.}
\footnotetext[3]{References~\onlinecite{BeckaPRSA63,BerlandJCP10}.}
\end{table*}

Table~\ref{table_MS} shows the results for the equilibrium lattice constant
$a_{0}$ and lattice energy $E_{\text{latt}}$ of the molecular solids NH$_{3}$
(ammonia), CO$_{2}$ (carbon dioxide),
and C$_{6}$H$_{12}$N$_{4}$ (hexamethylenetetramine).
$E_{\text{latt}}$ is defined as the difference between the total energy
(per molecule) of the crystal and the total energy of one isolated molecule.
These three systems, which have a cubic cell, are members of the X23 test set\cite{ReillyJCP13}
of molecular solids which is an improvement of the C21 test
set.\cite{OterodelaRozaJCP12} The C21 and X23 sets have been used in a certain
number of studies for testing functionals.
\cite{OterodelaRozaJCP12,ReillyJCP13,ReillyJPCL13,BuckoJCTC13,CarterJCTC14,KronikACR14,MoellmannJPCC14,HojaWCMS16,BrandenburgPRB16,BrandenburgPCCP16,CutiniJCTC16,HermannJCTC18,ThomasJCTC18,DolgonosJPCA18,MortazaviJPCL18,LobodaJCP18,BrandenburgJCP18,LeBlancJCTC18}
From our results, we can see that most dispersion-corrected functionals except vdW-DF lead
to reasonably small errors for the equilibrium lattice constant $a_{0}$.
Compared to the other dispersion-corrected
functionals, SCAN+rVV10 has a more pronounced tendency to underestimate $a_{0}$ ($-3\%$ for NH$_{3}$
and $-2\%$ CO$_{2}$), while SCAN (and TM) without NL-vdW correction was already pretty good
and compete with the best dispersion-corrected functionals.
We can see that also PBEsol+rVV10s leads to a large underestimation ($-4\%$) for NH$_{3}$.
However, we note that the experimental values for $a_{0}$ are not corrected for the
zero-point vibration effect, which, as mentioned in Ref.~\onlinecite{ReillyJCP13},
may increase the lattice constant by 1\%. Thus, a slight understimation in $a_{0}$
should be expected.

For the lattice energy $E_{\text{latt}}$, the most accurate functionals are
rev-vdW-DF2 and revPBE-D3(BJ), which lead to errors below 10\% for the three
systems. However, vdW-DF-cx and vdW-DF2 are also rather accurate, while
the most inaccurate functionals are optB88-vdW
(see also Ref.~\onlinecite{LobodaJCP18}), PBEsol+rVV10, and optB86b-vdW
that show errors above 20\% for all three molecular solids.
Note that, curiously, PBEsol+rVV10 leads to an overestimation for NH$_{3}$, but
to an underestimation for CO$_{2}$ and C$_{6}$H$_{12}$N$_{4}$.

For this test set of molecular solids, the functional that is overall
the most accurate is rev-vdW-DF2. Actually, rev-vdW-DF2 is the most accurate
for the lattice constant and the lattice energy. However, in order to be
fair, in particular since our test consists of only three systems, we should
also mention that other functionals, like revPBE-D3(BJ), vdW-DF-cx, or
vdW-DF2 seem to be pretty accurate overall. The plain MGGAs SCAN and TM lead
to large errors only for the lattice energy of C$_{6}$H$_{12}$N$_{4}$.

Concerning other dispersion-corrected DFT methods, a recent collection of
results from the literature for the full X23 test set can be found in Loboda
\textit{et al}.\cite{LobodaJCP18} Methods which should be of similar accuracy
as the best NL-vdW functionals are for instance
B86bPBE+XDM\cite{BeckeJCP05,OterodelaRozaJCP12,Johnson17} and
PBE+MBD,\cite{TkatchenkoPRL09,TkatchenkoPRL12,AmbrosettiJCP14,HermannJCTC18}
which are both atom-pairwise methods with density-dependent dispersion coefficients.

\subsection{\label{molecules}Molecules}

\begin{table*} 
\caption{\label{table_AE6}Atomization energy (in eV) for the molecules of the
AE6 test set. The units of the MRE and MARE are \%.
The reference results are from experiment.\cite{LynchJPCA03}
The functionals are separated into two groups, those which contain a dispersion
term (NL or atom-pairwise), and those which do not.
Within each group, the functionals are ordered by increasing MARE.
The errors (indicated in parenthesis) larger than 5\% are underlined.}
\begin{ruledtabular}
\begin{tabular}{lcccccccccc} 
Method & SiH$_{4}$             & SiO                   & S$_{2}$               & C$_{3}$H$_{4}$        & C$_{2}$H$_{2}$O$_{2}$ & C$_{4}$H$_{8}$        & ME & MAE & MRE & MARE \\ 
\hline
Without dispersion \\
SCAN\footnotemark[1]                 &    14.03 (0) &     8.02 (-4) &     4.73 (\underline{7}) &    30.48 (0) &    27.30 (-1) &    49.93 (0) &    -0.01 &     0.17 &      0.5 &      2.0 \\ 
TM\footnotemark[2]                   &    13.76 (-2) &     8.14 (-2) &     4.84 (\underline{10}) &    30.47 (0) &    27.69 (1) &    49.82 (0) &     0.02 &     0.20 &      1.1 &      2.5 \\ 
PBE\footnotemark[3]                  &    13.58 (-3) &     8.50 (2) &     4.98 (\underline{13}) &    31.27 (2) &    28.84 (5) &    50.64 (2) &     0.54 &     0.67 &      3.5 &      4.5 \\ 
PBEsol\footnotemark[3]               &    14.03 (0) &     8.90 (\underline{7}) &     5.36 (\underline{22}) &    32.51 (\underline{6}) &    30.27 (\underline{10}) &    52.85 (\underline{6}) &     1.56 &     1.56 &      8.6 &      8.6 \\ 
LDA\footnotemark[3]                  &    15.04 (\underline{8}) &     9.70 (\underline{16}) &     5.86 (\underline{33}) &    34.79 (\underline{14}) &    32.75 (\underline{19}) &    56.60 (\underline{14}) &     3.36 &     3.36 &     17.3 &     17.3 \\ 
With dispersion \\
vdW-DF2\footnotemark[3]              &    13.94 (0) &     8.19 (-2) &     4.31 (-2) &    29.98 (-2) &    27.06 (-1) &    48.46 (-3) &    -0.44 &     0.44 &     -1.7 &      1.7 \\ 
vdW-DF\footnotemark[3]               &    13.91 (-1) &     8.04 (-3) &     4.40 (0) &    29.80 (-3) &    26.94 (-2) &    48.53 (-3) &    -0.49 &     0.49 &     -1.9 &      1.9 \\ 
SCAN+rVV10\footnotemark[1]           &    14.04 (0) &     8.11 (-3) &     4.76 (\underline{8}) &    30.60 (0) &    27.61 (1) &    50.12 (1) &     0.11 &     0.18 &      1.2 &      2.0 \\ 
revPBE-D3(BJ)\footnotemark[3]        &    13.51 (-3) &     8.18 (-2) &     4.79 (\underline{9}) &    30.52 (0) &    27.90 (2) &    49.63 (0) &    -0.00 &     0.28 &      0.8 &      2.6 \\ 
rVV10\footnotemark[3]                &    13.52 (-3) &     8.43 (1) &     4.80 (\underline{9}) &    30.87 (1) &    28.31 (3) &    49.83 (0) &     0.20 &     0.35 &      1.8 &      2.9 \\ 
optB88-vdW\footnotemark[3]           &    14.11 (1) &     8.49 (2) &     4.85 (\underline{10}) &    30.94 (1) &    28.37 (3) &    50.43 (1) &     0.43 &     0.43 &      3.1 &      3.1 \\ 
vdW-DF-cx\footnotemark[3]            &    14.11 (1) &     8.46 (2) &     4.96 (\underline{13}) &    31.05 (2) &    28.56 (4) &    50.83 (2) &     0.57 &     0.57 &      3.8 &      3.8 \\ 
rev-vdW-DF2\footnotemark[3]          &    14.15 (1) &     8.56 (3) &     4.92 (\underline{12}) &    31.35 (3) &    28.82 (5) &    51.11 (3) &     0.72 &     0.72 &      4.3 &      4.3 \\ 
PBE-D3(BJ)\footnotemark[3]           &    13.63 (-3) &     8.52 (2) &     5.01 (\underline{14}) &    31.36 (3) &    28.93 (5) &    50.85 (2) &     0.62 &     0.74 &      3.9 &      4.7 \\ 
PBE+rVV10L\footnotemark[3]           &    13.61 (-3) &     8.53 (2) &     5.02 (\underline{14}) &    31.36 (3) &    28.95 (5) &    50.84 (2) &     0.62 &     0.75 &      3.9 &      4.8 \\ 
optB86b-vdW\footnotemark[1]          &    14.06 (1) &     8.68 (4) &     5.11 (\underline{16}) &    31.11 (2) &    28.92 (5) &    50.79 (2) &     0.68 &     0.68 &      4.9 &      4.9 \\ 
C09-vdW\footnotemark[3]              &    14.18 (1) &     8.62 (3) &     5.07 (\underline{15}) &    31.53 (3) &    29.10 (\underline{6}) &    51.56 (3) &     0.91 &     0.91 &      5.4 &      5.4 \\ 
PBEsol+rVV10s\footnotemark[3]        &    14.06 (1) &     8.92 (\underline{7}) &     5.39 (\underline{22}) &    32.59 (\underline{7}) &    30.36 (\underline{11}) &    53.01 (\underline{6}) &     1.62 &     1.62 &      8.9 &      8.9 \\ 
Reference                            &    13.98 &     8.33 &     4.41 &    30.56 &    27.46 &    49.83 \\ 
\end{tabular} 
\end{ruledtabular}
\footnotetext[1]{Calculated with \textsc{VASP}.}
\footnotetext[2]{Calculated with \textsc{deMon} non-self-consistently using PBE orbitals/density.}
\footnotetext[3]{Calculated with \textsc{CP2K}.}
\end{table*} 

All results presented so far were obtained for periodic solids, the focus
of the present work. However, as additional information we now
provide a snapshot of the accuracy of the functionals for finite systems
by considering the atomization energy of molecules.
Table~\ref{table_AE6} shows the results obtained for the AE6 test set
of six molecules.\cite{LynchJPCA03}

A well know problem at the GGA level of approximation is the difficulty
(and actually the quasi-impossibility) to get with the same functional
\textit{very} accurate results for the lattice constants and cohesive energies
of strongly bound solids and atomization energies of molecules.
In fact, even targeting only two of these three properties seems unachievable, and
the results in Refs.~\onlinecite{PerdewPRL08,ZhaoJCP08,PerdewPRL09,FabianoPRB10,HaasPRB11}
illustrate this problem for the lattice constant of solids and atomization
energy of molecules.
For this it is necessary to use functionals from higher rungs of Jacob's ladder,
MGGAs or hybrids, to get accurate results for both properties
simultaneously.\cite{PerdewPRL09,SunPRL15,TaoPRL16,DellaSalaIJQC16,SchimkaJCP11}
As seen in Table~\ref{table_AE6}, the dispersion-corrected GGA functionals
have the same problems as the GGA, which is expected since adding a dispersion
term to a functional should in principle have a rather limited effect on the results for
strongly bound systems (in particular if a dispersion term of small magnitude
like some of those of the rVV10-type is used, see Fig.~\ref{fig_kernel}). 
Indeed, the five most accurate GGA-based functionals
for the atomization energy (MAE below 0.5~eV), namely vdW-DF2, vdW-DF,
revPBE-D3(BJ), rVV10, and optB88-vdW are also the worst for the lattice constant $a_{0}$
(see Table~\ref{table_strong}). The reverse is also true: some of the most
accurate NL-vdW GGAs for $a_{0}$, e.g., C09-vdW or PBEsol+rVV10s
lead to the worst results for the AE6 atomization energy with a MAE that is several times
larger than for vdW-DF and vdW-DF2. However, note that vdW-DF-cx is rather well-balanced
since it is reasonably accurate for both the lattice constant and the molecular
atomization energy.

As mentioned, a dispersion term in the functional should be of relatively small
importance for covalently bound systems. Thus, as for the strongly bound solids
(Sec.~\ref{strongly}) some of the trends in the results correlate well
with the GGA enhancement factors $F_{\text{xc}}$ shown in Fig.~\ref{fig_Fxc}.
The factors $F_{\text{xc}}$ with the largest
magnitude (vdW-DF and vdW-DF2) lead to the best results, while a reduction
of the magnitude of $F_{\text{xc}}$ leads to more and more overbinding,
like PBEsol(+rVV10s) and ultimately LDA.

Thus, a GGA-based functional can not be among the best methods for more than
one of the three properties, which are the lattice constant and cohesive
energy of solids and the atomization energy of molecules.
MGGA functionals can alleviate this problem as exemplified by SCAN(+rVV10),
which belongs (more or less) to the most accurate functionals for
\textit{all three properties}. As shown in Table~\ref{table_AE6},
SCAN and SCAN+rVV10 (but also TM) lead to MAE below 0.2~eV and were competing
with the best GGA functionals for strongly bound solids (the only clear exceptions
are the GGAs optB88-vdW and rVV10 which are better for $E_{\text{coh}}$,
see Sec.~\ref{strongly}).

We mention that results for the AE6 molecules obtained with
several NL-vdW functionals were already available.\cite{CallsenPRB15}
Table~S10 of Ref.~\onlinecite{SM_NL-vdW} compares our results obtained with
two codes (\textsc{CP2K} and \textsc{VASP}) with those from
Ref.~\onlinecite{CallsenPRB15} obtained with \textsc{VASP}.
(To make the comparison possible, our vdW-DF2, optB88-vdW, and rev-vdW-DF2 results in Table~S10
were obtained using the non-spin-polarized version of the DRSLL and LMKLL kernels.)
The agreement between our two sets of results is in general very good,
which gives us confidence about the reliability of our results.
However, the agreement with the values from Ref.~\onlinecite{CallsenPRB15}
is good only in the case of PBE and rev-vdW-DF2 (except for C$_{4}$H$_{8}$
with the latter functional). In the case of vdW-DF2 and optB88-vdW extremely
large discrepancies are systematically obtained, the worst being for
SiH$_{4}$ with vdW-DF2 (8.9~eV from Ref.~\onlinecite{CallsenPRB15} and
14.2~eV in the present work with both codes).

\section{\label{summary}Discussion and conclusion}

A dozen of dispersion-corrected functionals have been tested on periodic solids
and the goal was to identify which of them are the most appropriate for solids.
In particular, the question is if there is a dispersion-corrected
functional that is reasonably accurate for all types of systems that
have been considered in the present work. The test set consisted of strongly
and weakly bound solids, and for the latter group three classes were considered:
rare gases, layered solids, and molecular solids. Additionally, results on a small
set of molecules were also shown.

\begin{figure}
\includegraphics[scale=0.56]{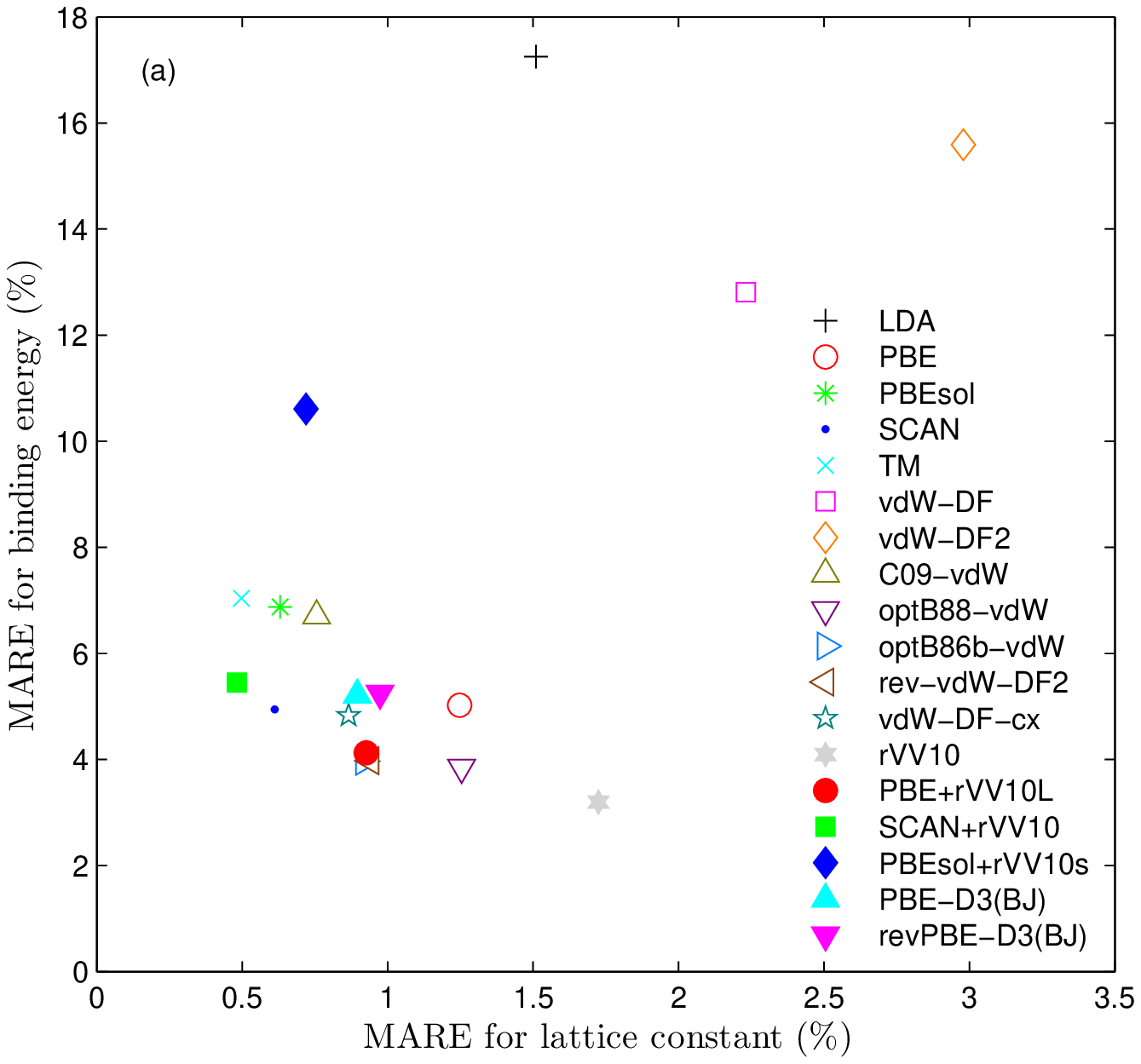}
\includegraphics[scale=0.56]{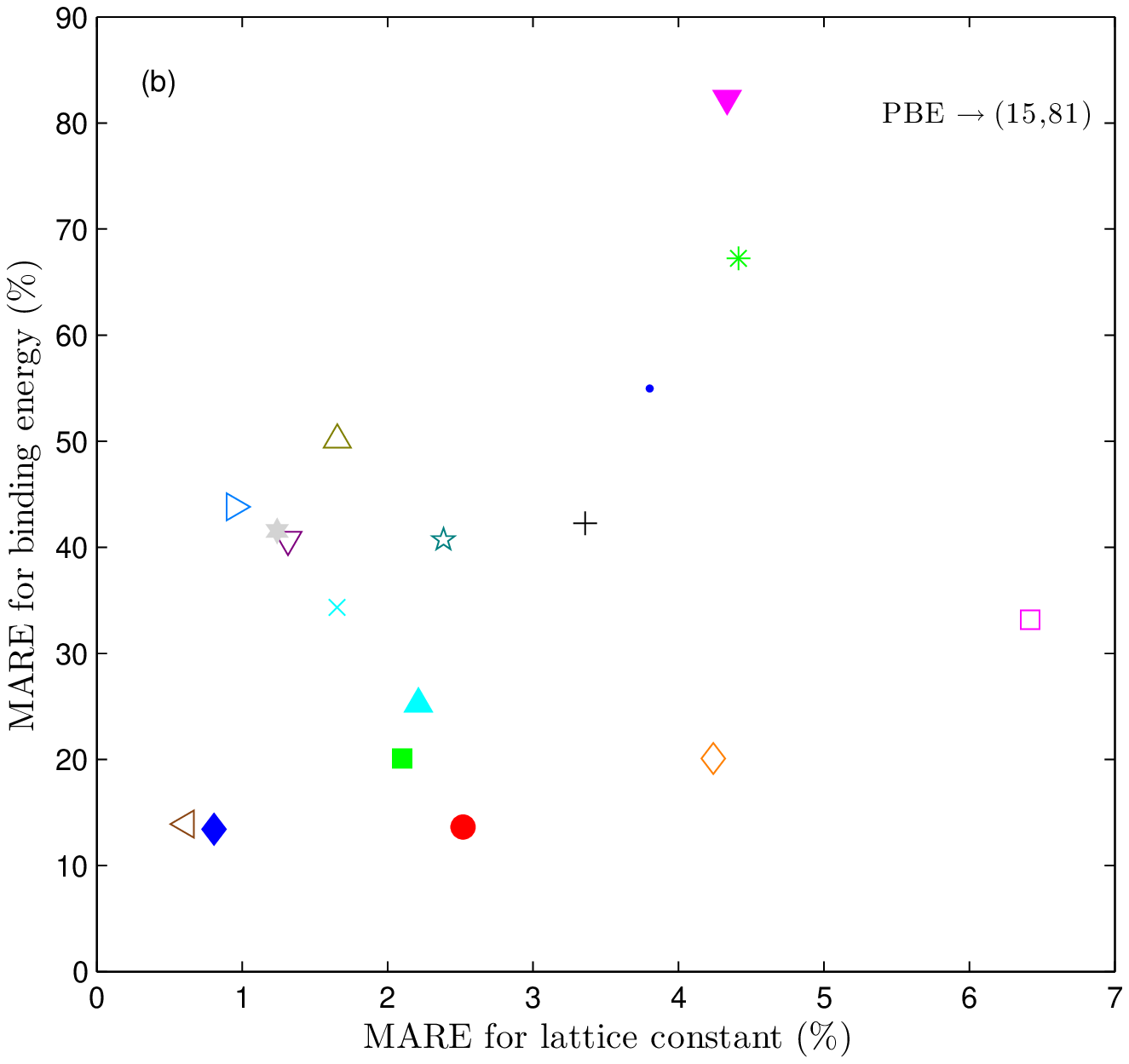}
\caption{\label{fig_MARE}MARE for lattice constant versus MARE for binding
energy for (a) the 44 strongly bound solids and (b) the 17 weakly bound solids
[($a_{0}$,$E_{\text{coh}}$) of the rare-gas solids, ($a_{0}$,$E_{\text{latt}}$)
of the molecular solids, and ($c_{0}$,$E_{b}$) of the layered solids].}
\end{figure}

Our results are summarized in Fig.~\ref{fig_MARE}.
For the strongly bound solids, the functionals that were considered as giving
satisfying results for all properties (lattice constant, bulk modulus, and cohesive
energy) are the MGGA SCAN and the NL-vdW SCAN+rVV10, PBE+rVV10L, optB86b-vdW,
rev-vdW-DF2, and vdW-DF-cx. The atom-pairwise methods PBE-D3(BJ) and revPBE-D3(BJ)
are also quite accurate.

In the case of the rare-gas solids, rev-vdW-DF2 and and PBE-D3(BJ)
are the most accurate overall (lattice constant and cohesive energy).
The results on the hexagonal layered solids have shown that only three functionals
provide reasonably small errors for all properties (intralayer and interlayer
lattice constants and interlayer binding energy) and for most solids:
PBE+rVV10L, SCAN+rVV10, and rev-vdW-DF2, however, none of them is
clearly superior to the two others.
Finally, for the molecular solids, rev-vdW-DF2 and revPBE-D3(BJ) lead overall to
the smallest errors for the lattice constant and cohesive energy.

From this summary the conclusion is the following. rev-vdW-DF2 is among
the most accurate methods for all three classes of weakly bound solids and is
therefore a recommended functional for treating weak interactions in solids.
Remarkably, rev-vdW-DF2 leads to no single catastrophic results, at least not
in our test set of systems with weak interactions. 
rev-vdW-DF2 does not belong to the list of the top-performing functionals
for strongly bound interactions, however the results are actually relatively
fair overall: although not among the best for the lattice constant
it is still better than PBE, and excellent for the cohesive energy.
Thus, overall rev-vdW-DF2 seems to be a very good compromise for
solid-state calculations and, furthermore, it is not based on a MGGA functional
but on a GGA, which leads to practical advantages. MGGA functionals lead to more expensive
calculations\cite{BienvenuJCTC18,MejiaRodriguezPRB18} and may require
denser grids for integrations as observed for
SCAN.\cite{BrandenburgPRB16,YaoJCP17}
However, the advantage of MGGA functionals is to be generally more accurate
as shown again in the present work for molecules.
It is worth to mention that very recently, Fischer \textit{et al}.
\cite{FischerJCP19} showed that rev-vdW-DF2 is one of the most accurate
functionals (among fourteen dispersion-corrected ones) for the structural
and energetic properties of a set of sixteen SiO$_{2}$ and AlPO$_{4}$ frameworks.
Thus, this consolidates the conclusion of the present work.

We finish by mentioning that the recently proposed PBEsol+rVV10s functional
shows mixed performances. While it is one of the most accurate for the lattice
constant of weakly bound solids [see Fig.~\ref{fig_MARE}(b)], it is not
recommended for the cohesive energy of strongly bound and molecular solids.

\begin{acknowledgments}

This work was supported by projects F41 (SFB ViCoM) and P27738-N28 of the
Austrian Science Fund (FWF) and by the TU-D doctoral college (TU Wien).
Part of this work was granted access to the HPC resources of
[TGCC/CINES/IDRIS] under allocation 2017-A0010907682 made by GENCI.
We are grateful to Ferenc Karsai for help regarding \textsc{VASP} calculations.

\end{acknowledgments}

\bibliography{references}

\end{document}